\documentclass[ALICE,manyauthors]{cernphprep}
\usepackage[comma,square,numbers,sort&compress]{natbib}
\usepackage{hyperref}
\usepackage{graphicx}  
\usepackage{dcolumn}   
\usepackage{bm}        
\usepackage{amssymb}   
\usepackage{amsfonts}
\usepackage{graphics}
\usepackage{grffile}   
\usepackage{epsfig}
\usepackage{units}
\usepackage[usenames]{color}
\usepackage[normalem]{ulem} 
\usepackage{color}
\usepackage[utf8]{inputenc}
\usepackage[T1]{fontenc}
\usepackage{subfigure}

\usepackage{rotating}
\usepackage{draftcopy}
\usepackage{grffile}
\usepackage{url}
\usepackage{epstopdf}
\usepackage{subfigure}
\usepackage[usenames,dvipsnames,svgnames,table]{xcolor}
\DeclareGraphicsExtensions{.pdf, .png, .jpg, .ps, .eps, .gif}
\usepackage{rotating}
\usepackage{enumerate}
\usepackage{comment}
\usepackage{indentfirst}
\usepackage{rotating}
\usepackage[left]{lineno}
\usepackage[titletoc]{appendix}
\usepackage{amsfonts}
\usepackage{esint}
\usepackage{amssymb}
\usepackage{epigraph}
\usepackage{textcomp}
\usepackage{color}
\usepackage[usenames,dvipsnames,svgnames,table]{xcolor}
\usepackage{booktabs}
\usepackage{microtype}
\usepackage{verbatim}
\usepackage{hyperref}
\usepackage{upgreek}
\usepackage{mathtools}
\newcommand{\sqs}{\mbox{$\sqrt{s_{\rm NN}}$}\xspace}

\makeatletter
\newcommand{\thickhline}{%
    \noalign {\ifnum 0=`}\fi \hrule height 1pt
    \futurelet \reserved@a \@xhline
}
\newcolumntype{"}{@{\hskip\tabcolsep\vrule width 1pt\hskip\tabcolsep}}
\makeatother

\begin{document}%

\begin{titlepage}
\PHyear{2020}       
\PHnumber{126}      
\PHdate{8 July}  

\title{Pion--kaon femtoscopy and the lifetime of the hadronic phase in Pb$-$Pb collisions at $\bm{\sqrt{{\textit s}_{\rm NN}}}\bm{=}\mathbf{2.76}$ TeV}
\ShortTitle{Pion--kaon femtoscopy in Pb$-$Pb collisions at $\mathbf{\sqrt{{\textit s}_{\rm NN}}}=2.76$ TeV}

\Collaboration{ALICE Collaboration%
         \thanks{See Appendix~\ref{app:collab} for the list of collaboration members}}
\ShortAuthor{ALICE Collaboration}       
\begin{abstract}

In this paper, the first femtoscopic analysis of pion--kaon correlations at the LHC is reported. The analysis
was performed on the Pb--Pb collision data at $\mathbf{\sqrt{{\textit s}_{\rm NN}}} =2.76$ TeV recorded with the ALICE detector.
The non-identical particle correlations probe the spatio-temporal separation between sources of different particle species as well as the average source size of the emitting system.  
The sizes of the pion and kaon sources increase with centrality, and pions are emitted closer to the centre of the system and/or later than kaons. This is naturally expected in a system with strong radial flow and is qualitatively reproduced by hydrodynamic models.  
ALICE data on pion--kaon emission asymmetry are consistent with (3+1)-dimensional viscous hydrodynamics coupled to a statistical hadronisation model, resonance propagation, and decay code THERMINATOR 2 calculation, with an additional time delay between 1 and 2~fm$/c$ for kaons. 
The delay can be interpreted as evidence for a significant hadronic rescattering phase in heavy-ion collisions at the LHC.

\end{abstract}
\end{titlepage}

\setcounter{page}{2} 

\section{Introduction}
The main goal of the heavy-ion programme at the Large Hadron Collider (LHC) is to study the deconfined state of strongly interacting matter. This state, where the relevant degrees of freedom are quarks and gluons, is called the quark-gluon plasma (QGP). Experimental results from RHIC suggest that the QGP behaves as a fluid with small specific viscosity~\cite{Arsene:2004fa,Back:2004je,Adams:2005dq,Adcox:2004mh}. The characteristics in momentum space can be accessed from radial and elliptic flow, transverse momentum spectra or from event-by-event fluctuations. The space-time structure, relevant for the size and pressure gradients of the system, can be accessed using two-particle correlations. 

Non-identical particle correlations are sensitive to the relative space-time emission shifts of different particle species~\cite{Lednicky:1995vk, Voloshin:1997jh,Kisiel:2009eh}. The difference between mean emission space-time coordinates of two particle species at freeze-out is called emission asymmetry. 
It occurs as a consequence of the collective expansion of the system, the presence of short-lived resonances decaying into the considered particles, the radial flow of these resonances,  
and the possibility of having additional rescattering between the chemical and kinetic boundaries of the evolution of the system~\cite{Kisiel:2009eh}.
Measurements of correlations of non-identical particles 
in low-energy heavy-ion collisions allowed one to establish an emission time ordering of the nuclear fragments~\cite{Kotte:1999gr,Gourio:2000tn}. 
In relativistic heavy-ion collisions they provided independent evidence of collective transverse expansion in Au--Au collisions at $\sqrt{s_{\rm NN}}=130$~GeV at the Relativistic Heavy Ion Collider (RHIC)~\cite{Adams:2003qa}.

The Hanbury Brown and Twiss (HBT)~\cite{Kopylov:1972qw,Lednicky:1981su,Adler:2001zd,Adams:2003ra,Lisa:2005dd,Adam:2015vna} pion correlation radii are a measure of the source size of pions of a given momentum. 
Together with measurements of the elliptic flow and the transverse momentum spectra of identified particles they have been fundamental in identifying the relevant stages of ultra-relativistic heavy-ion collisions and their properties~\cite{Broniowski:2008vp}. Furthermore, a recent measurement of the kaon femtoscopic radii in Pb--Pb collisions~\cite{Acharya:2017qtq} showed that (when compared for the same event centrality and pair $m_{\rm T}$) they are systematically larger than the ones from pions and those predicted by models based on a hydrodynamic evolution coupled to statistical hadronisation. 
Only after including the hadronic rescattering phase could the model \cite{Shapoval:2014wya} reproduce the data for pions and kaons simultaneously. The mean emission time of kaons (11.6~fm$/c$) and of pions (9.5~fm$/c$) were reported~\cite{Acharya:2017qtq}. The  difference is attributed to the rescattering through the K$^{*}$ resonance.

Particle yields and spectra add further support to models which include the formation of a dense hadronic phase in the final stages of the evolution of the fireball created in heavy-ion collisions.
The suppression or the enhancement of the yield (with respect to pp collisions) of short-lived resonances due to rescattering (suppression) or regeneration (enhancement) in the hadronic phase has been proposed as an observable for the estimation of the lifetime and properties of the hadronic phase \cite{Torrieri:2001ue,Bleicher_2003,Bellwied:2010pr}. The measurements of several resonances, from the very short-lived $\uprho$ meson ($\tau=1.4$~fm$/c$), K$^{*}$ ($\tau=4$~fm$/c$), $\Lambda(1520)$ ($\tau=10$~fm$/c$) to longer-lived $\upphi$ ($\tau=46$~fm$/c$), demonstrate strong suppression of short-lived resonances in central collisions  \cite{Acharya:2018qnp,Abelev:2014uua,ALICE:2018ewo}. The observed suppression can result from a long-lasting hadronic rescattering phase.

Recently, pion--kaon correlations were studied theoretically with a (3+1) viscous hydrodynamic model~\cite{Bozek:2011ua}, coupled to the statistical hadronisation, resonance decay, and propagation code THERMINATOR 2 [28].
The model uses a parameterisation of the equation of state interpolating between the lattice results ~\cite{Borsanyi:2010cj} for high temperatures and the hadron gas equation of state at low temperatures. 
The hadronisation occurs via the Cooper-Frye formalism without distinction  between chemical and kinetic freeze-out. 
No further interactions between the hadrons are considered, however, the emission time of each species can be delayed by hand, mimicking the effect of rescattering. The femtoscopic emission asymmetry was shown to be highly sensitive to this
delay.
Moreover, it can be decoupled from other mechanisms like flow or resonance contributions present at freeze-out, including the K$^{*}$ resonance~\cite{Kisiel:2018wie}. This approach has been explored for pion--kaon pairs. 
Detailed predictions for different emission scenarios for the pion--kaon radii and their emission asymmetry as a function of the source volume have been made for Pb--Pb collisions at $\sqrt{s_{\rm NN}}$ = 2.76 TeV in~\cite{Kisiel:2018wie}. 

In this work $\rm \uppi^{+}K^{+}$, $\rm \uppi^{-}K^{+}$, $\rm \uppi^+K^{-}$, and $\rm \uppi^{-}K^{-}$ momentum correlations are analysed using the femtoscopy technique. 
Two methods are used to evaluate the emission asymmetry in order to strengthen the results. 
The first method decomposes the correlations into terms of one dimensional spherical harmonic (SH) coefficients~\cite{Kisiel:2009iw} while  the second one is based on the Cartesian representation of the correlation function~\cite{Lednicky:1995vk}. 
The source size parameter $R_{\rm out}$ and the emission asymmetry $\mu_{\rm out}$ are measured as a function of the cube root of the average charged-particle multiplicity density $\langle\text{d}N_{\rm ch}/\text{d}\eta\rangle^{1/3}$. Finally, the obtained results are compared with detailed model calculations~\cite{Kisiel:2018wie} assuming the previously found delayed kaon emission \cite{Acharya:2017qtq}.

\section{Data selection} \label{sec:data_selection}
In this paper, pion--kaon correlation results obtained with Pb–Pb collisions at $\sqs$ = 2.76 TeV are presented. 
This measurement used 40 million events collected by ALICE in 2011.
A detailed description of the ALICE detector and its performance in the LHC Run 1 (2009--2013) is given in \cite{Aamodt:2008zz,Abelev:2014ffa}.

Events were triggered and classified according to their centrality determined using the measured signal amplitudes in the V0 detectors \cite{Abelev:2013qoq}. 
Three trigger configurations were used: minimum bias, semi-central (10--50\% collision centrality), and central (0--10\% collision centrality)~\cite{Abelev:2013qoq}. 
The analyses were performed in six centrality classes: (0--5\%), (5--10\%), (10--20\%), (20--30\%), (30--40\%), and (40--50\%), separately for positive and negative magnetic field polarity.
The reconstructed primary vertex is required to lie
within $\pm 7$~cm of the nominal interaction point along the beam axis in order to have
uniform tracking and particle identification performance.

Charged particle tracking is 
performed using the Time Projection Chamber (TPC)~\cite{Aamodt:2008zz,Dellacasa:2000bm} and the Inner Tracking System (ITS)  \cite{Aamodt:2008zz}. 
The ITS allows for high spatial resolution in determining the primary collision vertex.  
In this analysis, the determination of the track momenta was performed using tracks reconstructed only from TPC signals and constrained to the primary vertex. 
A TPC track segment is reconstructed from at least 70 space points (clusters) out of a maximum of 159. 
The $\chi^{2}$ of the track fit, normalised to the number of degrees of freedom, is required to be $\chi^{2}/\text{ndf}<2$.
The distances of closest approach (DCA) of a track to the primary vertex in the transverse ($\rm DCA_{\text{xy}}$) and longitudinal ($\rm DCA_{z}$) directions are required to be less than 2.4 cm and 3.2 cm, respectively. 
These selections are imposed to reduce the contamination from  secondary tracks originating from weak decays and from interaction with the detector material.  
The transverse momenta and pseudorapidities of pions and kaons were restricted to  $0.19< p_{\text T} < 1.5$~GeV$/c$ and $|\eta| < 0.8$.
All selections are summarised in Table~\ref{tab:trackcuts}.

The  charged-particle tracks are identified as pions and kaons using the combined information of their specific ionisation energy loss (d$E/\text{d}x$) in the TPC and the time-of-flight information from the Time-Of-Flight (TOF) detectors \cite{particleidentification}. 
For each reconstructed particle, the signals from both the TPC and the TOF (d$E/\text{d}x$ and time of flight, respectively) are compared with the ones predicted for a pion or kaon. A value $N_{\sigma}$ is assigned to each track denoting the number of standard deviations between the measured track d$E$/d$x$ or time of flight and the expected one.
For pions, the signal (d$E/\text{d}x$ for $p_{\rm T} <500$~MeV$/c$, combined d$E/\text{d}x$ and time of flight above this value) is allowed to differ from the calculation by 3$\sigma$. 
For kaons, five selections were used, as detailed in Table~\ref{tab:trackcuts}, together with variations used for uncertainty estimation.
The selection criteria are optimised to obtain a high-purity sample while maximising efficiency, especially in the regions where
separating kaons from other particle species is challenging.
The purity was estimated from Monte Carlo simulations using the HIJING~\cite{Wang:1991hta} event generator coupled to the GEANT3~\cite{Brun:1994aa} transport package and was found to be above 98\% for both the pion and kaon samples.

\begin{table}[th!]
\centering
\caption{Single particle selection criteria, together with particle identification variations used for uncertainty estimation.}
\label{tab:trackcuts}
\begin{tabular}{l|l|l|l}
  \hline
  \multicolumn{4}{c}{Track selection} \\ \hline
    $p_{\rm T}$  & \multicolumn{3}{c}{$0.19<p_{\rm T}<1.5$ GeV/$c$} \\ \hline
   $|\eta|$ & \multicolumn{3}{c}{$< 0.8$ }\\ \hline
    $\rm DCA_{\rm transverse}$ to primary vertex & \multicolumn{3}{c}{$< 2.4$ cm} \\ \hline
  $\rm DCA_{\rm longitudinal}$ to primary vertex & \multicolumn{3}{c}{$< 3.0$ cm} \\ \hline  
      \hline
  \multicolumn{4}{c}{Kaon selection} \\ \hline
   & default & loose & strict \\ \hline
    $N_{\sigma,\rm TPC}$ (for $p < 0.4$~GeV/$c$) & $< 2$ & $< 2.5$ & $< 2$ \\ \hline
    $N_{\sigma,\rm TPC}$ (for $0.4 < p < 0.45$ GeV/$c$) & $< 1$ & $< 2$ & $< 1$ \\ \hline
    $N_{\sigma,\rm TPC}$ (for $p > 0.45$~GeV/$c$) & $< 3$ & $< 3$ & $< 2$ \\ \hline
    $N_{\sigma,\rm TOF}$ (for $0.5 < p < 0.8$ GeV/$c$) & $< 2$ & $< 3$ & $< 2$ \\ \hline
    $N_{\sigma,\rm TOF}$ (for $0.8 < p < 1.0$ GeV/$c$) & $< 1.5$ & $< 2.5$ & $< 1.5$ \\ \hline
    $N_{\sigma,\rm TOF}$ (for $1.0 < p < 1.5$ GeV/$c$) & $< 1$ & $< 2$ & $< 1$ \\ \hline
  \hline
  \multicolumn{4}{c}{Pion selection} \\ \hline
     & default & loose & strict \\ \hline
    $N_{\sigma,\rm TPC}$ (for $p < 0.5$~GeV/$c$) & $< 3$  & $< 3$  & $< 2.5$ \\ \hline
    $\sqrt{N_{\sigma,\rm TPC}^2+N_{\sigma,\rm TOF}^2}$ (for $p > 0.5$~GeV/$c$) & $< 3$  & $< 3$  & $< 2.5$ \\ \hline
    
\end{tabular}
\end{table}

The identified tracks from each event are combined into pairs. Two-particle detector acceptance effects, including track splitting, track merging, as well as effects coming from $\rm \gamma\rightarrow e^{+}e^{-}$ conversion, contribute to the measured distributions. 
The following selections are applied to reduce these effects.
For pairs of tracks within $|\Delta\eta| < 0.1$ an exclusion on the  fraction of merged points is introduced.
The merged fraction is defined as the ratio of the number of steps of $1$ cm considered in the TPC radius range for which the distance between the tracks is less than $3$ cm to the total number of steps. 
Pairs with a merged fraction above 3\% were removed. The $\rm e^{+}e^{-}$ pairs originating from photon conversions can be misidentified as a real pion--kaon pair and it is necessary to remove spurious correlations arising from such pairs.  
These pairs are removed if their invariant mass, assuming the electron mass for both particles, is less than 0.002 GeV$/c^2$, and  the relative polar angle, $\Delta\theta$, between the two tracks is less than 0.008 rad.

\section{Correlation functions}\label{sec:corr_func}

The femtoscopic correlation function $C({\bm{k}}^*)$, as a function of the pion and kaon relative three-momenta $\bm{k}^*=\frac{1}{2}(\bm{p}^*_\uppi-\bm{p}^*_\text{K})$ in the pair rest frame (PRF) indicated with the asterisk, is constructed as 
\begin{equation}
    C(\bm{k}^*) = \mathcal{N}\frac{A({\bm{k}}^*)}{B(\bm{k}^*)},\label{eq:femto_cf}
\end{equation}
where $A(\bm{k}^*)$ is the distribution  constructed from the same event and $B(\bm{k}^*)$ is the reference distribution from particles belonging to different events using the event mixing method~\cite{Kopylov:1974th}. 
The normalisation constant $\mathcal{N}$ is used to ensure that the ratio reaches unity outside the momentum range where the correlation function is affected by final state interactions, i.e.  $0.15<k^*<0.20$ GeV$/c$, where $k^*= \left|{\bm{k}}^*\right|$. The average transverse momentum of pions and kaons belonging to pairs with $k^*<40$~MeV/$c$ is $0.27$~GeV/$c$ (std. dev. $0.07$~GeV/$c$)  and $0.93$~GeV/$c$ (std. dev. $0.23$~GeV/$c$), respectively, independent of centrality.

The first and second moments of the distribution of the spatio-temporal separation of emission points in the PRF can be obtained from correlation functions either in the three-dimensional Cartesian representation~\cite{Lednicky:1995vk} or using its decomposition into spherical harmonics (SH)~\cite{Brown:2005ze,Kisiel:2009iw}.  
The three-momentum and position differences can be projected onto the out-side-long orthogonal axes, where the long axis is the beam axis, the out axis is in the direction of the transverse pair velocity in the laboratory system, while the side axis is perpendicular to the long and out axes~\cite{Pratt:1986cc,Bertsch:1989vn}.  At midrapidity, the emission asymmetry -- displacement between pion and kaon sources -- can exist only in the out direction~\cite{Kisiel:2018wie}. In this work, the emission asymmetry in the out direction is obtained with two different methods and they are explained hereafter.

The SH decomposition allows one to project the three-dimensional information contained in the correlation function into a set of one-dimensional distributions. The method applied here uses the direct decomposition of $A(\bm{k}^*)$ and $B(\bm{k}^*)$ during the filling of the discrete distributions~\cite{Kisiel:2009iw}. The numerator can be written as
\begin{equation}
    A(\bm{k}^*) = \sqrt{4\pi} \sum_{l=0}^{\infty}\sum_{m=0}^{l} A_{l}^{m}(k^*)Y_l^m(\theta^*,\varphi^*),
    \end{equation}
where $Y_l^m(\theta^*,\varphi^*)$ are the spherical harmonics and $A_{l}^{m}(k^*) = \frac{1}{4\pi}\int_{4\pi} A(\bm{k}^*){Y_l^m}^{*}(\theta^*,\varphi^*)\text{d} \Omega^*$. A similar definition is valid also for the denominator. The $l<3$ terms from the infinite set of numerator and denominator distributions 
are filled for each reconstructed pair using the corresponding $Y_l^m(\theta^*,\varphi^*)$ weight for its $\theta^*$ and $\varphi^*$ angles.
From these one-dimensional distributions, the components of the correlation function can be calculated following the method introduced in~\cite{Kisiel:2009iw}.

\begin{figure}[tbh!]
\begin{center}
\includegraphics[scale=0.6]{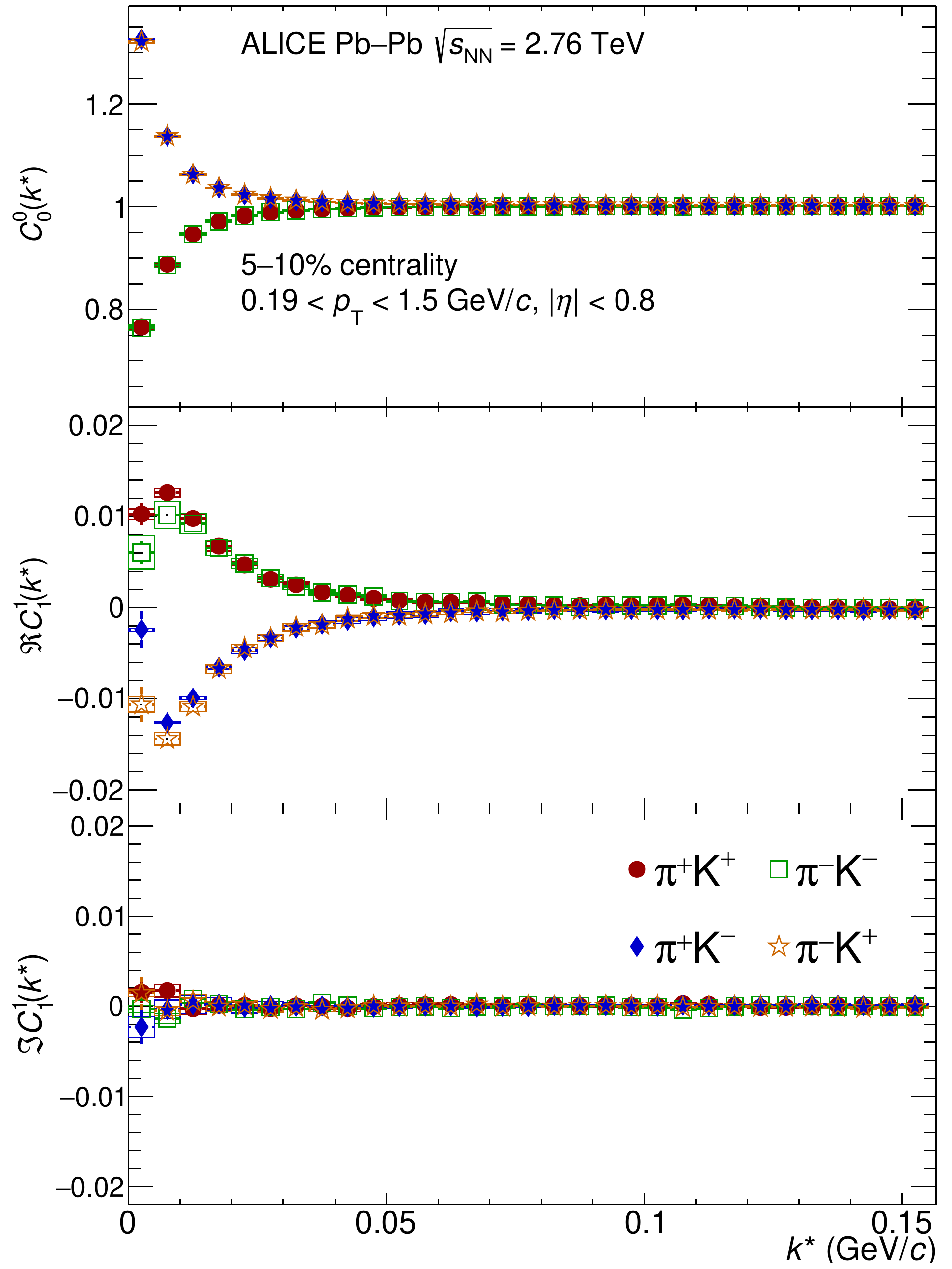}
\caption{The $C^{0}_{0}$ (top panel), $\Re C^{1}_{1}$ (middle panel), and $\Im C^{1}_{1}$ (bottom panel) SH components of the charged pion--kaon femtoscopic correlation functions for Pb--Pb collisions at \sqs = 2.76 TeV in the 5--10\% centrality class, positive field polarity. The different charge combinations of pions and kaons are shown with different colours and markers. 
The statistical and systematic uncertainties are shown as vertical bars and boxes, respectively.}
\label{fig1CorrFunSH}
\end{center}
\end{figure}

The femtoscopic information relevant for the emission asymmetry measurement is contained in two one-dimensional correlation functions, $C^{0}_{0}$ and the real part of $C^{1}_{1}$, where $C^{i}_{j}$ is defined as $A^{i}_{j}/B^{i}_{j}$.
The $C^{0}_{0}$  and $\Re C^{1}_{1}$ functions are mostly sensitive to the source size and the emission asymmetry, respectively~\cite{Kisiel:2009iw}.
Additionally, the values of $C^{0}_{1}$ (asymmetry in the long direction) and $\Im C^{1}_{1}$ are checked for zero emission asymmetry. 
Their deviations from zero may indicate track reconstruction problems in the detector. 
Higher order components are small and irrelevant for this analysis. 

\begin{figure}[tbh!]
\begin{center}
\includegraphics[scale=0.6]{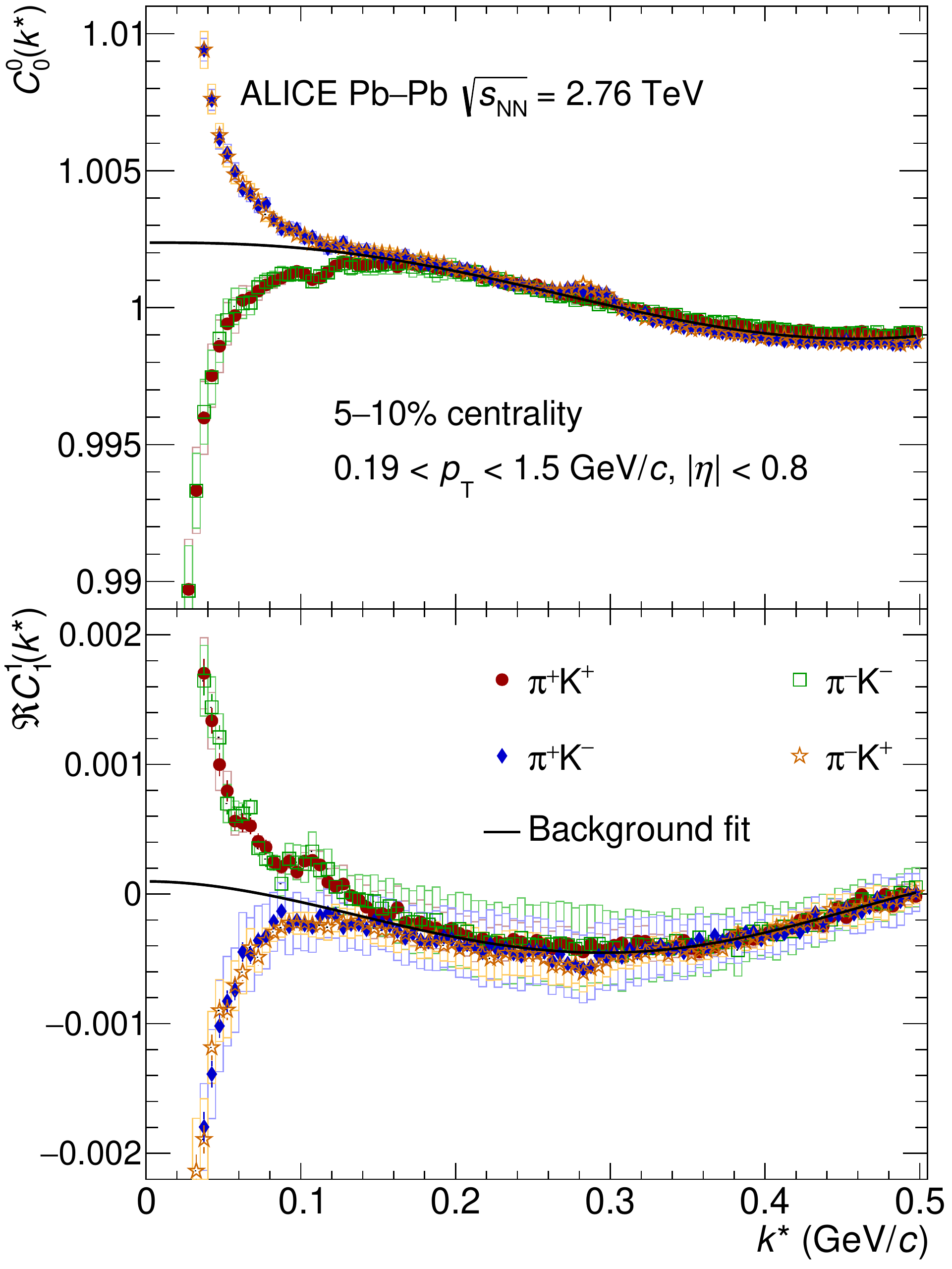}
\caption{The $C^{0}_{0}$ (top panel) and $\Re C^{1}_{1}$ (botton panel) components of the pion--kaon correlation functions in the 5--10\% centrality class showing the non-femtoscopic background in the spherical-harmonic representation, positive field polarity. The background fit corresponds to a 6$^{\text{th}}$ order polynomial function common for all charge combinations. The two structures visible in the correlation function at 0.11 GeV$/c$ and at 0.29 GeV$/c$  correspond to the remaining effect from track merging and the K$^{*}$ resonance, respectively. The statistical and systematic uncertainties are shown as vertical bars and boxes, respectively.}
\label{fig2Bkg}
\end{center}
\end{figure}

The $C^{0}_{0}$, $\Re C^{1}_{1}$, and $\Im C^{1}_{1}$ components of the correlation function in the SH representation are shown in Fig.~\ref{fig1CorrFunSH} for the different pairs. For like-sign pairs, the $C^{0}_{0}$ correlation goes below unity at low $k^*$, reflecting the repulsive character of the mutual Coulomb interaction. For unlike-sign pairs, the effect is opposite (see also Fig.~\ref{fig2Bkg}). For the $\Re C^{1}_{1}$ correlation function, the deviation from unity is directly related to the emission asymmetry between the two particle species. The $\Im C^{1}_{1}$ should be flat by symmetry and thus is a good check for detector and analysis biases.

For the Cartesian representation analysis, the reconstructed pairs were divided into two different correlation functions, namely $C_{+}({k^{*}})$ and $C_{-}({k^{*}})$, where the sign reflects the sign of $k^{*}_{\rm out}$. 
These correlation functions represent two different scenarios where the first particle (by construction the pion) is faster or slower than the second one (the kaon). The difference between them reflects the space-time emission asymmetry.

It can be observed from 
Fig.~\ref{fig2Bkg} that the correlation function is not exactly equal to unity at large values of $k^*$, but has some intrinsic slope mainly due to the presence of elliptic flow, resonance decays, and due to global conservation of energy and momentum. 
These background correlations have to be subtracted before fitting the correlation functions in both the SH and Cartesian representations. 
The procedure to estimate the non-femtoscopic background is described in detail in~\cite{Kisiel:2017gip}, where it is shown that for $\uppi^{\pm}K^{\pm}$ pairs the non-femtoscopic baseline can be parameterised by a common $6^{\rm th}$ order polynomial function for all pair combinations.
The same approach is used to correct the effect of non-femtoscopic background in the present analysis and the resulting background estimation is shown in Fig.~\ref{fig2Bkg} as a solid black line for the $C_0^0$ and $\Re C_1^1$ components of pion--kaon pairs of different charge sign combinations.

\section{Fitting of the correlation functions}\label{sec:fitting}

The experimental correlation functions in both representations are compared to theoretical functions calculated with the software package CorrFit~\cite{Kisiel:2004fcn}. These functions are calculated as   
\begin{equation}
    C(\bm{k}^*) = \frac{\int S({r^*}) |\Psi_{\uppi \text{K}}({r^*},\bm{k}^*)|^2 \text{d}^4 {r^*}} {\int S({r^*}) \text{d}^4 {r^*}},\label{eq:cf_theory}
\end{equation}
where the four-vector ${r^*}={x^*_{\uppi}}-{x^*_\text{K}}$ is the space-time position difference of a pion and a kaon, $S({r^*})$ is the source emission function which is the probability of emitting a pair of particles at a given position difference. 
The possible dependence of the source on $\bm{k}^*$ has been neglected. 
This approximation has been proven for radii larger than 1--2~fm~\cite{Lisa:2005dd}. 
$\Psi_{\uppi\text{K}}$ is the pion--kaon pair wave function. 
It accounts for the Coulomb and strong final-state interactions (FSI), the former  being dominant for the correlation effect~\cite{Kisiel:2018wie}.

In order to be able to compare the resulting radii to those obtained from identical-particle femtoscopy, we parameterise the source in the longitudinally comoving coordinate system (LCMS), defined for each pair such that the longitudinal pair momentum vanishes. The relative two-particle source can be expressed as
\begin{equation}
S(\bm{r}) \propto \exp\left(-\frac{[r_{\text{out}} -
    \mu_{\text{out}}]^{2}}{2R_{\text{out}}^{2}}
  -\frac{r_{\text{side}}^{2}}{2R_{\text{side}}^{2}}
  -\frac{r_{\text{long}}^{2}}{2R_{\text{long}}^{2}}
   \right),
\label{eq:Slcms}
\end{equation}
where $R_{\text{out}}$, $R_{\text{side}}$, and $R_{\text{long}}$ are the femtoscopic radii in the three directions and $\mu_{\text{out}}$ is the emission asymmetry. 
In order to avoid a large set of fitting parameters, the relations $R_{\text{side}} = R_{\text{out}}$ and $R_{\text{long}} = 1.3 R_{\text{out}}$ are used, which are based on measured radii from identical pion femtoscopy from the same experimental data~\cite{Adam:2015vna}.
In this approach only two independent parameters are needed to characterise the correlation function for the whole system: $\mu_{\text{out}}$ and $R_{\text{out}}$.
In order to (numerically) compute the fit function corresponding to Eq.~\ref{eq:cf_theory}, the relative positions between pions and kaons are sampled from Eq.~\ref{eq:Slcms}, while their momenta are sampled from the respective experimental distributions from the same data set. 
The positions and momenta are then boosted from the LCMS to the PRF. The fit value is the mean wave function squared in the PRF.

The fitting procedure also accounts for the purity of the sample, defined as the percentage of the properly identified primary particle pairs originating from the 3D Gaussian profile, referred to as the ``Gaussian core''. 
Products of decays of long lived resonances are considered as not correlated.
Following the method proposed in~\cite{Kisiel:2009eh}, the values for the purity parameter depend on the misidentification, 
on the secondary contamination from weak decays, and on the percentage of pions and kaons that come from strongly decaying resonances constituting the long-range tails in the source distribution, outside the Gaussian core. These three purity factors are denoted as $p$, $f$, and $g$, respectively. The pair purity (also referred to as the primary fraction) is evaluated independently for each centrality class and magnetic field polarity and is defined as:
\begin{equation}
	  P_{\uppi^{\pm}K^{\pm}} = p_{\uppi^{\pm}} \cdot p_{{\rm K}^{\pm}} \cdot f_{\uppi^{\pm}} \cdot f_{{\rm K}^{\pm}} \cdot g.
\end{equation}
All parameters except $g$ are obtained from a detailed simulation of the detector response calculated using the HIJING Monte Carlo model with particle transport performed by GEANT3. 
The $g$  values are taken from a calculation in~\cite{Kisiel:2009eh} following the methodology used in~\cite{Kisiel:2018wie}. 
The total value of the pair purity is 0.73 for the 0--5\% centrality class and decreases smoothly to 0.61 for the 40--50\% centrality class. 

The experimental finite momentum resolution has been incorporated in the fitting procedure. The ideal three-momenta of 20~000 randomly selected pairs from analysed data per $k^*$ bin used in the fitting routine were smeared by the momentum-dependent experimental momentum and angular resolutions. These were obtained from Monte Carlo simulations using a detailed description of the experimental set-up.

\begin{figure}[tbh!]
\begin{center}
\includegraphics[width=0.49\textwidth]{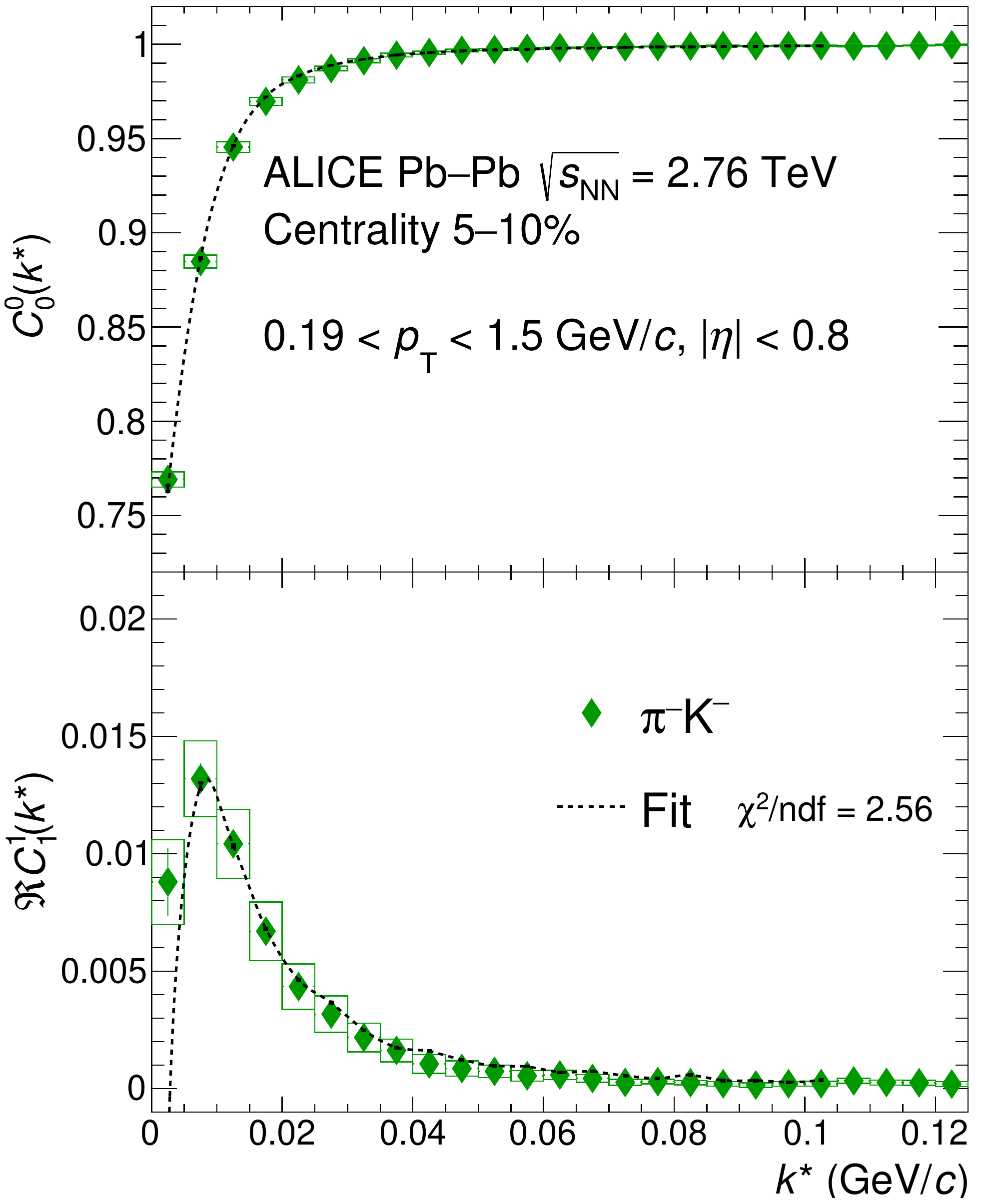}
\includegraphics[width=0.49\textwidth]{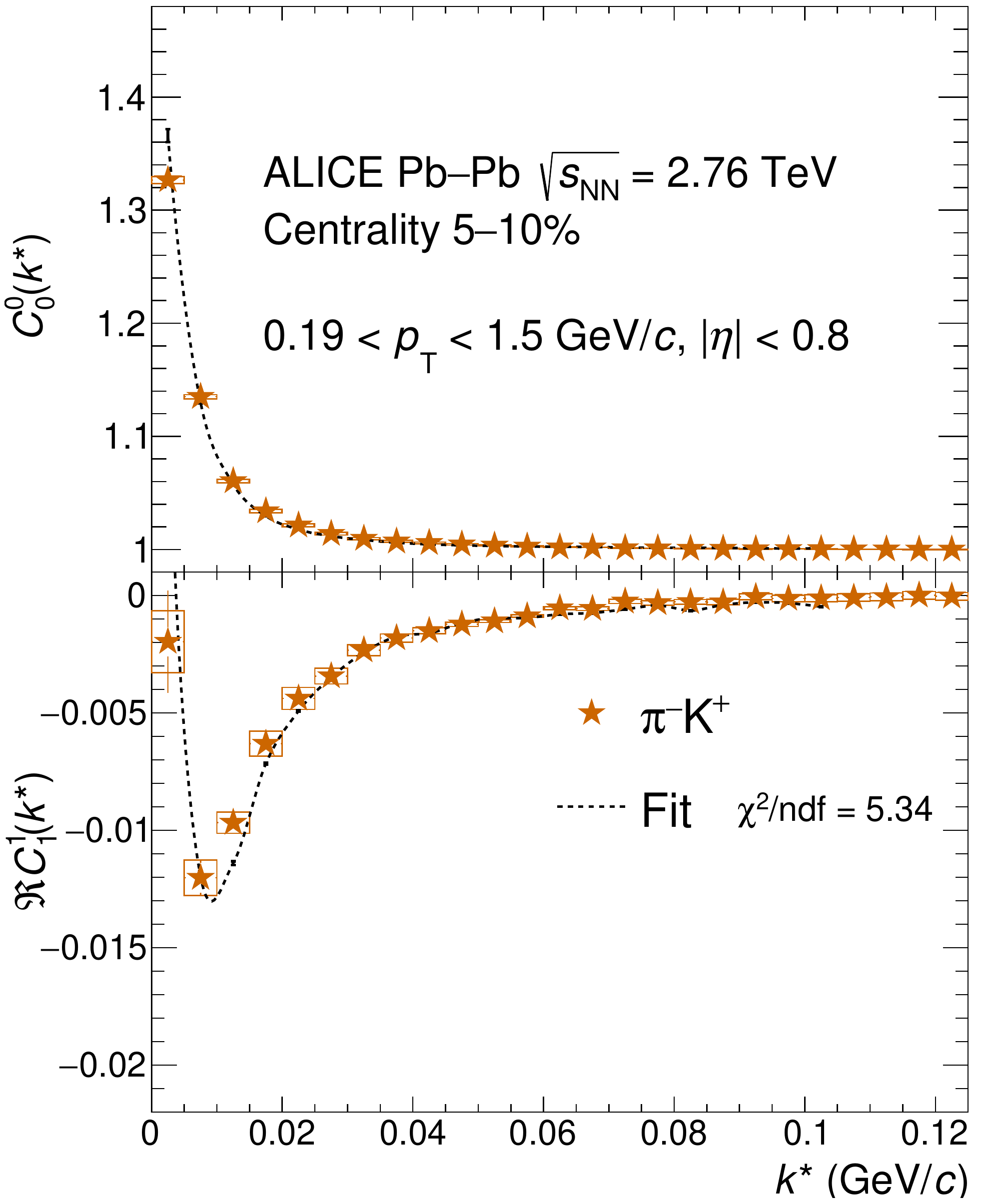}
\caption{The $C_0^0(k^*)$ and $\Re C_1^1(k^*)$ parts of the correlation function for (left) $\rm \uppi^{-}K^{-}$ and (right) $\rm \uppi^{-}K^{+}$ pairs, shown as markers for the 5--10\% centrality, with the  corresponding fits calculated using the CorrFit package shown as dashed lines. Only half of the statistics is used, corresponding to one magnetic field (positive field polarity). The statistical and systematic uncertainties are shown as vertical lines and boxes, respectively.}
\label{fig3_CorrWithFitSH}
\end{center}
\end{figure}

Each of the correlation functions obtained for the six event centralities, four charge combinations, and two polarities of the electric field have been fitted independently. The values of the radii and emission asymmetry are obtained using a $\chi^2$ minimisation in the $R_{\text{out}}-\mu_{\text{out}}$ plane.
The fitting is done in the range $0 < k^* < 0.1$ GeV$/c$ using the CorrFit package~\cite{Kisiel:2004fcn}. A fit example of the $C_0^0(k^*)$ and $\Re C_1^1(k^*)$ parts of the correlation function for $\uppi^{-}\text{K}^{-}$ and $\uppi^{-}\text{K}^{+}$ is shown in Fig.~\ref{fig3_CorrWithFitSH}. Note that the poor $\chi^2$ values reflect the residual deviations from a Gaussian distribution, rather than an improperly performed fit. The non-Gaussianity comes mainly from combining different pair transverse momenta, representing three spatial dimensions in a one-dimensional correlation function, and the presence of daughters of short-lived (up to $\omega$) resonance decays.

\begin{figure}[tbh!]
\begin{center}
\includegraphics[scale=0.8]{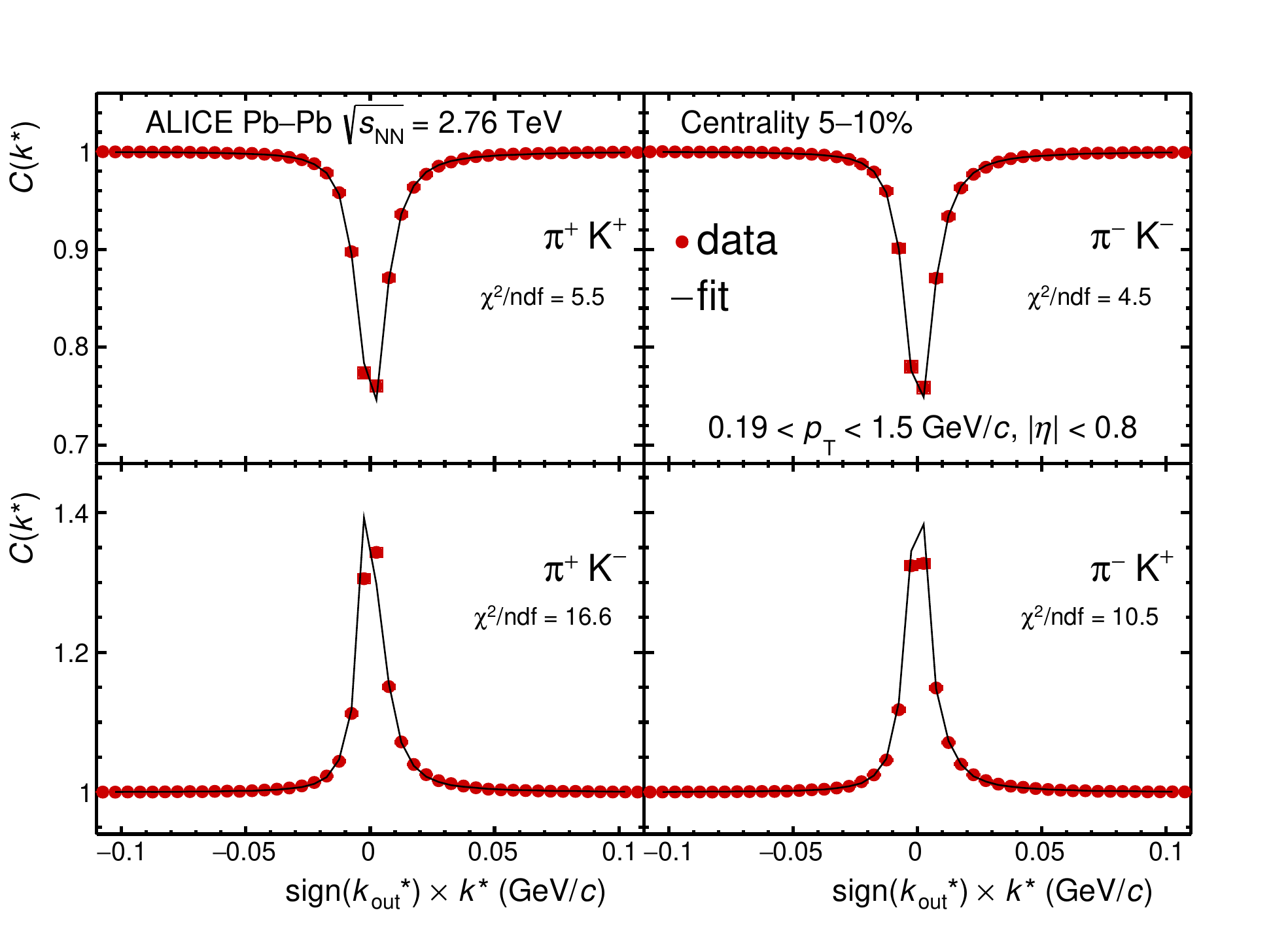}
\caption{Pion--kaon correlation functions in the Cartesian representation for all charge combinations. The $C_-$ is on the negative side of the $k^*$ axes while $C_+$ is on the positive. The femtoscopic fits are shown as a solid black line and were computed using the CorrFit package. The statistical and systematic uncertainties are smaller than the markers.}
\label{fig_CF_cartesian_with_fit}
\end{center}
\end{figure}

The systematic uncertainties are estimated by 
varying the particle identification and selection criteria, 
the normalisation range of the correlation functions, 
the background fit range of the polynomial that is used for estimation of non-femtoscopic contributions, 
the fit range, 
and the momentum resolution parameters used for smearing. 
Values of these variations and their individual contributions to the systematic uncertainty are summarised in Table~\ref{tab:syst}. All the systematic uncertainties are evaluated independently for each centrality class and the maximum value is reported in the table.
The primary pair fractions are treated separately. They introduce a significant and correlated systematic error for all centralities. 

The final uncertainty is obtained combining the systematic and statistical uncertainties using the covariance ellipses method.  
For each of the eight fit results (pair combinations and magnetic field polarities) as well as for each systematic variation, 10$^4$ points are generated following a two-dimensional Gaussian distribution in the $R_{\text{out}}$--$\mu_{\text{out}}$ space, where the mean and covariance are taken from the fit. 
The covariance ellipses are calculated from the sample of generated points in each centrality bin.
The systematic uncertainties used for the final result are  obtained  using 1$\sigma$ covariance ellipses. Negligible correlation between $R_{\text{out}}$--$\mu_{\text{out}}$ parameters is observed.

Additionally, the analysis was done in the Cartesian representation~\cite{Lednicky:1995vk} using the projected $C_{+}$ and $C_{-}$ correlation functions shown in Fig.~\ref{fig_CF_cartesian_with_fit}. 
The results of this analysis are fully compatible with those from SH within uncertainties. However, these results are not incorporated as another source of systematic uncertainty since the Cartesian method yields three times larger statistical uncertainties of $\mu_{\rm out}$.

\begin{table}[htbp!]
\caption{Input parameters to CorrFit used to fit the correlation functions and variation of relevant parameters and ranges used for the evaluation of the systematic uncertainties of $R_{\text{out}}$ and $\mu_{\rm out}$. The first three uncertainty sources affect the correlation functions and are visualised in Figs.~\ref{fig1CorrFunSH} and~\ref{fig2Bkg}. The uncertainties were estimated for all the centrality ranges independently and maximum value is reported. The variation of primary pair fractions was not included in the covariance ellipse calculation and is shown separately as a correlated model-dependent systematic uncertainty indicated with a $\dagger$ symbol. Uncertainties from fits using only Coulomb interaction, indicated with symbol $\ddagger$, are not included in the final systematic uncertainty. The ranges indicated with $^\S$ symbol include exclusion of 0.1--0.125 GeV/$c$ and 0.265--0.315 GeV/$c$, to account for splitting effects and $\rm K^*$ resonance. }
\centering
\begin{tabular}{@{} p{4cm}"p{2.5cm}|p{3.3cm}|p{1.6cm}|p{1.6cm} @{}}
\thickhline
{\bf Uncertainty source} & Default value & Variations & max $R_{\rm out}$ (\%) & max $\mu_{\rm out}$ (\%) \\
\thickhline
PID & Default in Table~\ref{tab:trackcuts} & Loose and strict in Table~\ref{tab:trackcuts} & 3.0 & 12.0 \\
\hline
Background fit range ($k^*$ in GeV/$c$) & 0.0--0.5$^\S$ & 0.0--0.265$^\S$, \hspace{1.5cm}0.125--0.5$^\S$ & 2.6 & 17.3 \\ 
\hline 
Normalisation range ($k^*$ in GeV/$c$) & 0.15--0.2 &  0.1--0.12, 0.18--0.21 &  3.3 & 18.0 \\ 
\hline
Fit range ($k^*$ in GeV/$c$) & 0--0.1 & 0--0.08/0.12, \hspace{1.5cm}0.005--0.1   & 3.7  & 13.4 \\  
\hline
Momentum resolution & Procedure from \cite{Aamodt:2008zz,Abelev:2014ffa} & $+$12\% &  3.6 & 10.3 \\ 
\thickhline
Primary fraction$\dagger$ & In Sec.~\ref{sec:fitting}& $\pm$10\% & 15.0$\dagger$ & 20.0$\dagger$\\ 
\hline
Analysis type & SH & Cartesian coordinates & 1.6 & 3.1 \\
\hline
$\Psi_{\uppi \text{K}}\ddagger$  & Strong and Coulomb & Coulomb only & 33.0$\ddagger$ & 8.7$\ddagger$ \\ 
\thickhline
\end{tabular}
  \label{tab:syst}
\end{table}

Fits to correlation functions considering only Coulomb interaction show a systematic and centrality-dependent decrease for $R_{\text{out}}$ of the order of 33\% with a significantly increased $\chi^2$ of the fit. For this reason these are not included in the evaluation of the uncertainties. However, the effect on the asymmetry parameter, supporting the prediction made in~\cite{Kisiel:2018wie}, is about 9\%, in line with other variations and demonstrating the prevalence of the Coulomb interaction for the emission asymmetry measurement. 

\section{Results}

\begin{figure}[htb!]
\begin{center}
\includegraphics[scale=0.8]{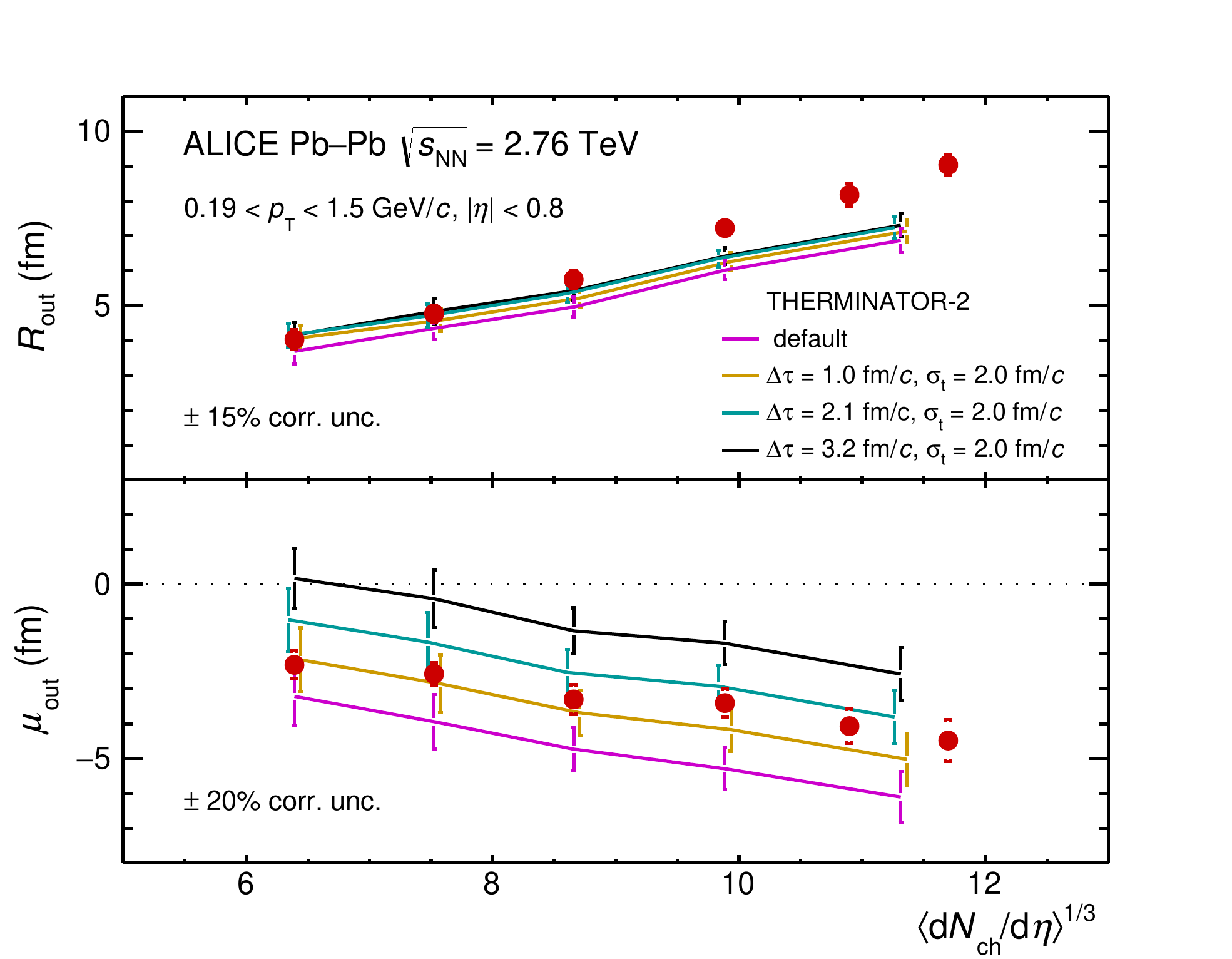}
\caption{Pion--kaon source size (upper panel) and emission asymmetry (lower panel) for Pb--Pb collisions at $\sqrt{s_{\rm NN}}$ = 2.76 TeV as a function of $\langle$d$N_{\rm ch}/$d$\eta\rangle^{1/3}$. The solid lines show predictions from calculation of source size and emission asymmetry using the THERMINATOR 2 model with default and selected values of additional delay with a mean time of $\Delta\uptau$ and width $\upsigma_{\text t}$ for kaons~\cite{Kisiel:2018wie}. The statistical and systematic uncertainties are combined and shown as square brackets. The uncertainty related to the fraction of primary pairs is reported separately as a correlated model-dependent systematic uncertainty of $\pm$15\% (20\%). }\label{fig:final_result}
\end{center}
\end{figure}

The final extracted radii, $R_{\text{out}}$, and emission asymmetry, $\mu_{\text{out}}$, are calculated as a weighted averages between the values obtained from the analysis of correlation functions corresponding to two magnetic field polarities and four possible charge combinations of charged pion--kaon pairs, using the SH representation.
The obtained values are shown as a function of $\langle$d$N_{\rm ch}/$d$\eta\rangle^{1/3}$ in Fig.~\ref{fig:final_result}. 
The radius increases smoothly from 4~fm to 9~fm when going from the 40--50\% centrality interval to 0--5\%.
At the same time, the emission asymmetry evolves from a starting value of $\mu_{\rm out}=-2.5$~fm and reaches $\mu_{\rm out}=-4$~fm for the most central events. 
In the same figure, the predictions published in~\cite{Kisiel:2018wie} are shown as lines for different hypotheses of the extra delay for kaons, starting from the default setting with no delay to a maximum of 3.2 fm$/c$ extra emission time. 
This delay reduces the asymmetry produced naturally which originates from the collective behaviour of the expanding system created in the collisions modelled with THERMINATOR 2~\cite{Kisiel:2005hn}. 
The agreement between the measured and predicted radii is good for peripheral events but measurements are larger by 1.5~fm for the most central events. 
On the other hand, the emission asymmetry measurement follows the predicted trends for all centralities. 
The data points lie between the curves corresponding to time delays of 1.0 and 2.1 fm$/c$. 

The model-dependent systematic errors of 15\% and 20\% for the radii and asymmetry, respectively, are present also in the theoretical prediction, as the same values for the fraction of particles within the Gaussian core are used to obtain the radii and emission asymmetry \cite{Kisiel:2009eh}. 
Therefore, this additional systematic uncertainty would synchronously move the results up and down  and the prediction lines without changing their interpretation. 

\section{Discussion}

In this work the first femtoscopy analysis of pion--kaon pairs at the LHC is presented. The collective behaviour of the matter created in Pb--Pb collisions generates a natural asymmetry in the emission of pions and kaons due to their different masses. 
This is related to the kaon emission distribution, which is more strongly influenced by flow than pions~\cite{Kisiel:2009eh}.
The analysis was implemented using the spherical harmonics and the Cartesian representation of the femtoscopic correlation function. 
The non-femtoscopic background present in the raw ratios was subtracted using a combined fit to the four possible charge combinations. 
The final results are compared to state-of-the-art hydrodynamical calculations where an additional delay for kaons was introduced to mimic the behaviour during the hadron rescattering phase.

The radii values predicted by the theoretical calculation~\cite{Kisiel:2018wie} have several assumptions included in the particle distributions which are different from the experiment. 
One of them is that the presence of the strong interaction does not modify the emission asymmetry  visible in the correlation functions. 
Our analysis confirms this statement; removal of strong interaction from the fit has significant influence on the radii (33\%) but moderate influence on the emission asymmetry  (9\%). 
Even though pions and kaons have been selected according to ALICE acceptance and momentum ranges, the optimisation of the purity of the data sample modified the transverse momentum distribution. 
This experimental effect biases the distributions towards lower momentum values, hence it increases the source radii.

The obtained width of the relative pion--kaon source, $R_{\text{out}}$, can be compared to the pion and kaon source radii extracted from identical-particle correlation analyses added in quadrature. 
The pion--kaon pairs used in the current analysis are predominantly composed of soft pions ($0.2\le m_{\text{T}}\le0.3$~GeV/$c$) and hard kaons ($1.0\le m_{\text{T}}\le1.3$~GeV/$c$). 
The pion and kaon source radii measured for these ranges of transverse mass ($m_{\rm T}$) in 0--10\% central collisions were 7--8.5~fm and 4--5~fm, respectively~\cite{Acharya:2017qtq}. 
Added in quadrature, this yields 8--10~fm, well in agreement with the most central pion--kaon point in Fig.~\ref{fig:final_result}. 
Similarly, for 30--50\% centrality class, the pion and kaon sources are 4--4.5~fm and 2--3~fm, respectively, and their combination yields 4.5--5.5~fm, again in reasonable agreement with the average of two most peripheral intervals in Fig.~\ref{fig:final_result}.

The emission asymmetry presented here coincides with the predictions calculated including a delay of the kaon emission of 1.0--2.1 fm$/c$. The difference between the $\mu_{\rm out}$ values predicted in Ref.~\cite{Kisiel:2018wie} and the measured value, averaged over centrality and normalised to the total uncertainty of our measurement, is shown in Table~\ref{tab:nsigmasTHERM}.
\begin{table}[htbp!]
\caption{Centrality-averaged difference between the $\mu_{\rm out}$ predicted using THERMINATOR with different values of the added kaon delay $\Delta\tau$~\cite{Kisiel:2018wie} and the one measured in this analysis, divided by the total uncertainty of the measurement $\sigma^{\rm exp}$.}
\centering
\begin{tabular}{c|r}
\thickhline
\noalign{\vskip 0.25mm}
 $\Delta\tau$ & $(\mu_{\rm out}^{\rm THERM} - \mu_{\rm out}^{\rm exp})/\sigma^{\rm exp}$ \\
\thickhline
 no delay  & $-$3.62~~~~~~~~~~~~~~\\
 \hline
 1.0~fm/$c$ & $-$1.02~~~~~~~~~~~~~~\\
  \hline
  2.1~fm/$c$ &  2.15~~~~~~~~~~~~~~\\
    \hline
 3.2~fm/$c$ & 5.26~~~~~~~~~~~~~~\\
\thickhline
\end{tabular}
  \label{tab:nsigmasTHERM}
\end{table}

The values obtained for the emission asymmetry are in line with those predicted by the hydrokinetic model \cite{Shapoval:2014wya}, the broken $m_{\text{T}}$ scaling of the radii of kaons with respect to pions observed in~\cite{Acharya:2017qtq}, and from the short-lived resonances measured by ALICE~\cite{Acharya:2018qnp,Abelev:2014uua,ALICE:2018ewo}. 
This measurement is another confirmation of the hadron rescattering phase.

In order to better understand the relevant effects influencing the emission asymmetry, it would be natural to continue the studies measuring other systems. 
It would be especially interesting to measure the $\uppi$p and Kp systems and probe the validity of the relation  $\mu_{\text{out}}^{\uppi\text{p}} = \mu_{\text{out}}^{\uppi\text{K}}+\mu_{\text{out}}^{\text{Kp}}$ \cite{Kisiel:2009eh}. Final-state interactions such as the ones taking place in a long-lasting rescattering phase might modify or distort this picture.

In summary, the first measurement of the emission asymmetry of pions and kaons for different centralities at the LHC has been performed.
$R_{\text{out}}$ was measured to be 9~fm for central collisions and decreases as a function of centrality to 4.5~fm for more peripheral collisions. 
At the same time, the magnitude of the emission asymmetry changed from $\mu_{\rm out}=-4.5$~fm to $\mu_{\rm out}=-2$~fm. 
This confirms the importance of the collective expansion of the system with the pions emitted closer to the centre of the collision and/or later than kaons. 
However, the collective motion is not enough to reproduce the trend of the emission asymmetry  which according to state-of-the-art models based on 3+1 viscous hydrodynamics demands an additional time delay of 1--2 fm$/c$ for kaons in order to reproduce the measured trend. 
This observation is in agreement with a hydrodynamic evolution of the expanding system and favors a stronger radial flow in central collisions together with a dense and long-lasting hadronic rescattering phase at the end of the evolution of the fireball at LHC energies.

\newenvironment{acknowledgement}{\relax}{\relax}
\begin{acknowledgement}
\section*{Acknowledgements}

The ALICE Collaboration would like to thank all its engineers and technicians for their invaluable contributions to the construction of the experiment and the CERN accelerator teams for the outstanding performance of the LHC complex.
The ALICE Collaboration gratefully acknowledges the resources and support provided by all Grid centres and the Worldwide LHC Computing Grid (WLCG) collaboration.
The ALICE Collaboration acknowledges the following funding agencies for their support in building and running the ALICE detector:
A. I. Alikhanyan National Science Laboratory (Yerevan Physics Institute) Foundation (ANSL), State Committee of Science and World Federation of Scientists (WFS), Armenia;
Austrian Academy of Sciences, Austrian Science Fund (FWF): [M 2467-N36] and Nationalstiftung f\"{u}r Forschung, Technologie und Entwicklung, Austria;
Ministry of Communications and High Technologies, National Nuclear Research Center, Azerbaijan;
Conselho Nacional de Desenvolvimento Cient\'{\i}fico e Tecnol\'{o}gico (CNPq), Financiadora de Estudos e Projetos (Finep), Funda\c{c}\~{a}o de Amparo \`{a} Pesquisa do Estado de S\~{a}o Paulo (FAPESP) and Universidade Federal do Rio Grande do Sul (UFRGS), Brazil;
Ministry of Education of China (MOEC) , Ministry of Science \& Technology of China (MSTC) and National Natural Science Foundation of China (NSFC), China;
Ministry of Science and Education and Croatian Science Foundation, Croatia;
Centro de Aplicaciones Tecnol\'{o}gicas y Desarrollo Nuclear (CEADEN), Cubaenerg\'{\i}a, Cuba;
Ministry of Education, Youth and Sports of the Czech Republic, Czech Republic;
The Danish Council for Independent Research | Natural Sciences, the VILLUM FONDEN and Danish National Research Foundation (DNRF), Denmark;
Helsinki Institute of Physics (HIP), Finland;
Commissariat \`{a} l'Energie Atomique (CEA) and Institut National de Physique Nucl\'{e}aire et de Physique des Particules (IN2P3) and Centre National de la Recherche Scientifique (CNRS), France;
Bundesministerium f\"{u}r Bildung und Forschung (BMBF) and GSI Helmholtzzentrum f\"{u}r Schwerionenforschung GmbH, Germany;
General Secretariat for Research and Technology, Ministry of Education, Research and Religions, Greece;
National Research, Development and Innovation Office, Hungary;
Department of Atomic Energy Government of India (DAE), Department of Science and Technology, Government of India (DST), University Grants Commission, Government of India (UGC) and Council of Scientific and Industrial Research (CSIR), India;
Indonesian Institute of Science, Indonesia;
Centro Fermi - Museo Storico della Fisica e Centro Studi e Ricerche Enrico Fermi and Istituto Nazionale di Fisica Nucleare (INFN), Italy;
Institute for Innovative Science and Technology , Nagasaki Institute of Applied Science (IIST), Japanese Ministry of Education, Culture, Sports, Science and Technology (MEXT) and Japan Society for the Promotion of Science (JSPS) KAKENHI, Japan;
Consejo Nacional de Ciencia (CONACYT) y Tecnolog\'{i}a, through Fondo de Cooperaci\'{o}n Internacional en Ciencia y Tecnolog\'{i}a (FONCICYT) and Direcci\'{o}n General de Asuntos del Personal Academico (DGAPA), Mexico;
Nederlandse Organisatie voor Wetenschappelijk Onderzoek (NWO), Netherlands;
The Research Council of Norway, Norway;
Commission on Science and Technology for Sustainable Development in the South (COMSATS), Pakistan;
Pontificia Universidad Cat\'{o}lica del Per\'{u}, Peru;
Ministry of Science and Higher Education, National Science Centre and WUT ID-UB, Poland;
Korea Institute of Science and Technology Information and National Research Foundation of Korea (NRF), Republic of Korea;
Ministry of Education and Scientific Research, Institute of Atomic Physics and Ministry of Research and Innovation and Institute of Atomic Physics, Romania;
Joint Institute for Nuclear Research (JINR), Ministry of Education and Science of the Russian Federation, National Research Centre Kurchatov Institute, Russian Science Foundation and Russian Foundation for Basic Research, Russia;
Ministry of Education, Science, Research and Sport of the Slovak Republic, Slovakia;
National Research Foundation of South Africa, South Africa;
Swedish Research Council (VR) and Knut \& Alice Wallenberg Foundation (KAW), Sweden;
European Organization for Nuclear Research, Switzerland;
Suranaree University of Technology (SUT), National Science and Technology Development Agency (NSDTA) and Office of the Higher Education Commission under NRU project of Thailand, Thailand;
Turkish Atomic Energy Agency (TAEK), Turkey;
National Academy of  Sciences of Ukraine, Ukraine;
Science and Technology Facilities Council (STFC), United Kingdom;
National Science Foundation of the United States of America (NSF) and United States Department of Energy, Office of Nuclear Physics (DOE NP), United States of America.    
\end{acknowledgement}
%
\bibliographystyle{utphys}
\bibliography{bibliography}

\providecommand{\href}[2]{#2}\begingroup\raggedright\begin{thebibliography}{10}

\bibitem{Arsene:2004fa}
{\bfseries BRAHMS} Collaboration, I.~Arsene {\em et~al.}, ``{Quark gluon plasma
  and color glass condensate at RHIC? The Perspective from the BRAHMS
  experiment},'' \href{http://dx.doi.org/10.1016/j.nuclphysa.2005.02.130}{{\em
  Nucl. Phys. A} {\bfseries 757} (2005) 1--27},
  \href{http://arxiv.org/abs/nucl-ex/0410020}{{\ttfamily
  arXiv:nucl-ex/0410020}}.

\bibitem{Back:2004je}
{\bfseries PHOBOS} Collaboration, B.~Back {\em et~al.}, ``{The PHOBOS
  perspective on discoveries at RHIC},''
  \href{http://dx.doi.org/10.1016/j.nuclphysa.2005.03.084}{{\em Nucl. Phys. A}
  {\bfseries 757} (2005) 28--101},
  \href{http://arxiv.org/abs/nucl-ex/0410022}{{\ttfamily
  arXiv:nucl-ex/0410022}}.

\bibitem{Adams:2005dq}
{\bfseries STAR} Collaboration, J.~Adams {\em et~al.}, ``{Experimental and
  theoretical challenges in the search for the quark gluon plasma: The STAR
  Collaboration's critical assessment of the evidence from RHIC collisions},''
  \href{http://dx.doi.org/10.1016/j.nuclphysa.2005.03.085}{{\em Nucl. Phys. A}
  {\bfseries 757} (2005) 102--183},
  \href{http://arxiv.org/abs/nucl-ex/0501009}{{\ttfamily
  arXiv:nucl-ex/0501009}}.

\bibitem{Adcox:2004mh}
{\bfseries PHENIX} Collaboration, K.~Adcox {\em et~al.}, ``{Formation of dense
  partonic matter in relativistic nucleus-nucleus collisions at RHIC:
  Experimental evaluation by the PHENIX collaboration},''
  \href{http://dx.doi.org/10.1016/j.nuclphysa.2005.03.086}{{\em Nucl. Phys. A}
  {\bfseries 757} (2005) 184--283},
  \href{http://arxiv.org/abs/nucl-ex/0410003}{{\ttfamily
  arXiv:nucl-ex/0410003}}.

\bibitem{Lednicky:1995vk}
R.~Lednick\'y, V.~L. Lyuboshits, B.~Erazmus, and D.~Nouais, ``How to measure
  which sort of particles was emitted earlier and which later,''
{\em Phys. Lett. B} {\bfseries 373} (1996) 30--34.

\bibitem{Voloshin:1997jh}
S.~Voloshin, R.~Lednicky, S.~Panitkin, and N.~Xu, ``{Relative space-time
  asymmetries in pion and nucleon production in noncentral nucleus-nucleus
  collisions at high-energies},''
  \href{http://dx.doi.org/10.1103/PhysRevLett.79.4766}{{\em Phys. Rev. Lett.}
  {\bfseries 79} (1997) 4766--4769},
  \href{http://arxiv.org/abs/nucl-th/9708044}{{\ttfamily
  arXiv:nucl-th/9708044}}.

\bibitem{Kisiel:2009eh}
A.~Kisiel, ``{Non-identical particle femtoscopy at {$\sqrt{s_{{\text
  {NN}}}}=200$}~GeV in hydrodynamics with statistical hadronization},''
  \href{http://dx.doi.org/10.1103/PhysRevC.81.064906}{{\em Phys. Rev. C}
  {\bfseries 81} (2010) 064906},
\href{http://arxiv.org/abs/0909.5349}{{\ttfamily arXiv:0909.5349 [nucl-th]}}.

\bibitem{Kotte:1999gr}
{\bfseries FOPI} Collaboration, R.~Kotte {\em et~al.}, ``{On the space-time
  difference of proton and composite particle emission in central heavy-ion
  reactions at 400 $A\cdot$MeV},''
  \href{http://dx.doi.org/10.1007/s100500050333}{{\em Eur. Phys. J. A}
  {\bfseries 6} (1999) 185--195},
\href{http://arxiv.org/abs/nucl-ex/9904007}{{\ttfamily arXiv:nucl-ex/9904007}}.

\bibitem{Gourio:2000tn}
{\bfseries INDRA} Collaboration, D.~Gourio {\em et~al.}, ``{Emission time scale
  of light particles in the system Xe+Sn at 50$A$ MeV. A probe for dynamical
  emission ?},'' \href{http://dx.doi.org/10.1007/PL00013604}{{\em Eur. Phys. J.
  A} {\bfseries 7} (2000) 245--253},
\href{http://arxiv.org/abs/nucl-ex/0001004}{{\ttfamily arXiv:nucl-ex/0001004}}.

\bibitem{Adams:2003qa}
{\bfseries STAR} Collaboration, J.~Adams {\em et~al.}, ``{Pion-kaon
  correlations in Au+Au collisions at {$\sqrt{s_{\text {NN}}}=130$}~GeV},''
  \href{http://dx.doi.org/10.1103/PhysRevLett.91.262302}{{\em Phys. Rev. Lett.}
  {\bfseries 91} (2003) 262302},
\href{http://arxiv.org/abs/nucl-ex/0307025}{{\ttfamily arXiv:nucl-ex/0307025
  [nucl-ex]}}.

\bibitem{Kopylov:1972qw}
G.~I. Kopylov and M.~I. Podgoretsky, ``Correlations of identical particles
  emitted by highly excited nuclei,''
{\em Sov. J. Nucl. Phys.} {\bfseries 15} (1972) 219--223.

\bibitem{Lednicky:1981su}
R.~Lednick\'y and V.~L. Lyuboshits, ``{Final State Interaction Effect on
  Pairing Correlations Between Particles with Small Relative Momenta},'' {\em
  Sov. J. Nucl. Phys.} {\bfseries 35} (1982) 770.
[Yad. Fiz.35,1316(1981)].

\bibitem{Adler:2001zd}
{\bfseries STAR} Collaboration, C.~Adler {\em et~al.}, ``{Pion interferometry
  of {$\sqrt{s_{\text {NN}}}=130$}~GeV Au+Au collisions at RHIC},''
  \href{http://dx.doi.org/10.1103/PhysRevLett.87.082301}{{\em Phys. Rev. Lett.}
  {\bfseries 87} (2001) 082301},
\href{http://arxiv.org/abs/nucl-ex/0107008}{{\ttfamily arXiv:nucl-ex/0107008
  [nucl-ex]}}.

\bibitem{Adams:2003ra}
{\bfseries STAR} Collaboration, J.~Adams {\em et~al.}, ``{Azimuthally sensitive
  HBT in Au + Au collisions at {$\sqrt{s_{\text {NN}}}=200$}~GeV},''
  \href{http://dx.doi.org/10.1103/PhysRevLett.93.012301}{{\em Phys. Rev. Lett.}
  {\bfseries 93} (2004) 012301},
\href{http://arxiv.org/abs/nucl-ex/0312009}{{\ttfamily arXiv:nucl-ex/0312009
  [nucl-ex]}}.

\bibitem{Lisa:2005dd}
M.~Lisa, S.~Pratt, R.~Soltz, and U.~Wiedemann, ``Femtoscopy in relativistic
  heavy ion collisions,'' {\em Ann. Rev. Nucl. Part. Sci.} {\bfseries 55}
  (2005) 311,
\href{http://arxiv.org/abs/nucl-ex/0505014}{{\ttfamily nucl-ex/0505014}}.

\bibitem{Adam:2015vna}
{\bfseries ALICE} Collaboration, J.~Adam {\em et~al.}, ``{Centrality dependence
  of pion freeze-out radii in Pb--Pb collisions at {$\sqrt{s_{\text
  {NN}}}=$}~2.76~TeV},''
  \href{http://dx.doi.org/10.1103/PhysRevC.93.024905}{{\em Phys. Rev. C}
  {\bfseries 93} no.~2, (2016) 024905},
\href{http://arxiv.org/abs/1507.06842}{{\ttfamily arXiv:1507.06842 [nucl-ex]}}.

\bibitem{Broniowski:2008vp}
W.~Broniowski, M.~Chojnacki, W.~Florkowski, and A.~Kisiel, ``{Uniform
  Description of Soft Observables in Heavy-Ion Collisions at {$\sqrt{s_{\text
  {NN}}}=200$}~GeV},''
  \href{http://dx.doi.org/10.1103/PhysRevLett.101.022301}{{\em Phys. Rev.
  Lett.} {\bfseries 101} (2008) 022301},
\href{http://arxiv.org/abs/0801.4361}{{\ttfamily arXiv:0801.4361 [nucl-th]}}.

\bibitem{Acharya:2017qtq}
{\bfseries ALICE} Collaboration, S.~Acharya {\em et~al.}, ``{Kaon femtoscopy in
  Pb--Pb collisions at $\sqrt{s_{\rm{NN}}}$ = 2.76 TeV},''
  \href{http://dx.doi.org/10.1103/PhysRevC.96.064613}{{\em Phys. Rev. C}
  {\bfseries 96} no.~6, (2017) 064613},
\href{http://arxiv.org/abs/1709.01731}{{\ttfamily arXiv:1709.01731 [nucl-ex]}}.

\bibitem{Shapoval:2014wya}
V.~M. Shapoval, P.~Braun-Munzinger, I.~A. Karpenko, and {\relax Yu}.~M.
  Sinyukov, ``{Femtoscopy correlations of kaons in Pb+Pb collisions at LHC
  within hydrokinetic model},''
  \href{http://dx.doi.org/10.1016/j.nuclphysa.2014.05.003}{{\em Nucl. Phys. A}
  {\bfseries 929} (2014) 1--8},
\href{http://arxiv.org/abs/1404.4501}{{\ttfamily arXiv:1404.4501 [hep-ph]}}.

\bibitem{Torrieri:2001ue}
G.~Torrieri and J.~Rafelski, ``{Strange hadron resonances as a signature of
  freezeout dynamics},''
  \href{http://dx.doi.org/10.1016/S0370-2693(01)00492-0}{{\em Phys. Lett. B}
  {\bfseries 509} (2001) 239--245},
\href{http://arxiv.org/abs/hep-ph/0103149}{{\ttfamily arXiv:hep-ph/0103149
  [hep-ph]}}.

\bibitem{Bleicher_2003}
M.~Bleicher and H.~Stoecker, ``Dynamics and freeze-out of hadron resonances at
  {RHIC},'' \href{http://dx.doi.org/10.1088/0954-3899/30/1/010}{{\em Journal of
  Physics G: Nuclear and Particle Physics} {\bfseries 30} no.~1, (Dec, 2003)
  S111--S118}. \url{https://doi.org/10.1088\%2F0954-3899\%2F30\%2F1\%2F010}.

\bibitem{Bellwied:2010pr}
R.~Bellwied and C.~Markert, ``{In-medium hadronization in the deconfined matter
  at {RHIC} and {LHC}},''
  \href{http://dx.doi.org/10.1016/j.physletb.2010.06.028}{{\em Phys. Lett. B}
  {\bfseries 691} (2010) 208--213},
\href{http://arxiv.org/abs/1005.5416}{{\ttfamily arXiv:1005.5416 [nucl-th]}}.

\bibitem{Acharya:2018qnp}
{\bfseries ALICE} Collaboration, S.~Acharya {\em et~al.}, ``{Production of the
  {$\rho$(770)${^{0}}$} meson in $pp$ and Pb--Pb collisions at
  {$\sqrt{s_{{\text {NN}}}}=2.76$} TeV},''
  \href{http://dx.doi.org/10.1103/PhysRevC.99.064901}{{\em Phys. Rev. C}
  {\bfseries 99} no.~6, (2019) 064901},
\href{http://arxiv.org/abs/1805.04365}{{\ttfamily arXiv:1805.04365 [nucl-ex]}}.

\bibitem{Abelev:2014uua}
{\bfseries ALICE} Collaboration, B.~B. Abelev {\em et~al.}, ``{$K^*(892)^0$ and
  $\phi(1020)$ production in Pb-Pb collisions at $\sqrt{s_{\rm NN}}$ = 2.76
  TeV},'' \href{http://dx.doi.org/10.1103/PhysRevC.91.024609}{{\em Phys. Rev.
  C} {\bfseries 91} (2015) 024609},
  \href{http://arxiv.org/abs/1404.0495}{{\ttfamily arXiv:1404.0495 [nucl-ex]}}.

\bibitem{ALICE:2018ewo}
{\bfseries ALICE} Collaboration, S.~Acharya {\em et~al.}, ``{Suppression of
  $\Lambda(1520)$ resonance production in central Pb--Pb collisions at
  $\sqrt{s_{\text{NN}}}$ = 2.76 TeV},''
  \href{http://dx.doi.org/10.1103/PhysRevC.99.024905}{{\em Phys. Rev. C}
  {\bfseries 99} (2019) 024905},
\href{http://arxiv.org/abs/1805.04361}{{\ttfamily arXiv:1805.04361 [nucl-ex]}}.

\bibitem{Bozek:2011ua}
P.~Bozek, ``{Flow and interferometry in 3+1 dimensional viscous
  hydrodynamics},'' \href{http://dx.doi.org/10.1103/PhysRevC.85.034901}{{\em
  Phys. Rev.} {\bfseries C85} (2012) 034901},
\href{http://arxiv.org/abs/1110.6742}{{\ttfamily arXiv:1110.6742 [nucl-th]}}.

\bibitem{Borsanyi:2010cj}
S.~Borsanyi, G.~Endrodi, Z.~Fodor, A.~Jakovac, S.~D. Katz, S.~Krieg, C.~Ratti,
  and K.~K. Szabo, ``{The QCD equation of state with dynamical quarks},''
  \href{http://dx.doi.org/10.1007/JHEP11(2010)077}{{\em JHEP} {\bfseries 11}
  (2010) 077}, \href{http://arxiv.org/abs/1007.2580}{{\ttfamily arXiv:1007.2580
  [hep-lat]}}.

\bibitem{Kisiel:2018wie}
A.~Kisiel, ``{Pion-kaon femtoscopy in Pb--Pb collisions at $\sqrt{s_{\rm
  NN}}=2.76$ TeV modeled in (3+1)D hydrodynamics coupled to Therminator 2 and
  the effect of delayed kaon emission},''
  \href{http://dx.doi.org/10.1103/PhysRevC.98.044909}{{\em Phys. Rev. C}
  {\bfseries 98} no.~4, (2018) 044909},
\href{http://arxiv.org/abs/1804.06781}{{\ttfamily arXiv:1804.06781 [nucl-th]}}.

\bibitem{Kisiel:2009iw}
A.~Kisiel and D.~A. Brown, ``{Efficient and robust calculation of femtoscopic
  correlation functions in spherical harmonics directly from the raw pairs
  measured in heavy-ion collisions},''
  \href{http://dx.doi.org/10.1103/PhysRevC.80.064911}{{\em Phys. Rev. C}
  {\bfseries 80} (2009) 064911},
\href{http://arxiv.org/abs/0901.3527}{{\ttfamily arXiv:0901.3527 [nucl-th]}}.

\bibitem{Aamodt:2008zz}
{\bfseries ALICE} Collaboration, K.~Aamodt {\em et~al.}, ``{The ALICE
  experiment at the CERN LHC},''
\href{http://dx.doi.org/10.1088/1748-0221/3/08/S08002}{{\em JINST} {\bfseries
  3} (2008) S08002}.

\bibitem{Abelev:2014ffa}
{\bfseries ALICE} Collaboration, B.~Abelev {\em et~al.}, ``{Performance of the
  ALICE Experiment at the CERN LHC},''
  \href{http://dx.doi.org/10.1142/S0217751X14300440}{{\em Int. J. Mod. Phys. A}
  {\bfseries 29} (2014) 1430044},
\href{http://arxiv.org/abs/1402.4476}{{\ttfamily arXiv:1402.4476 [nucl-ex]}}.

\bibitem{Abelev:2013qoq}
{\bfseries ALICE} Collaboration, B.~Abelev {\em et~al.}, ``{Centrality
  determination of Pb--Pb collisions at {$\sqrt{s_{\text{NN}}}=$}~2.76~TeV with
  ALICE},'' \href{http://dx.doi.org/10.1103/PhysRevC.88.044909}{{\em Phys. Rev.
  C} {\bfseries 88} no.~4, (2013) 044909},
\href{http://arxiv.org/abs/1301.4361}{{\ttfamily arXiv:1301.4361 [nucl-ex]}}.

\bibitem{Dellacasa:2000bm}
{\bfseries ALICE} Collaboration, G.~Dellacasa {\em et~al.},
``{ALICE: Technical design report of the time projection chamber},''.

\bibitem{particleidentification}
{\bfseries ALICE} Collaboration, K.~Aamodt {\em et~al.}, ``{Production of
  pions, kaons and protons in pp collisions at $\sqrt{s}= 900$ GeV with ALICE
  at the LHC},'' \href{http://dx.doi.org/10.1140/epjc/s10052-011-1655-9}{{\em
  Eur. Phys. J. C} {\bfseries 71} (2011) 1655},
\href{http://arxiv.org/abs/1101.4110}{{\ttfamily arXiv:1101.4110 [hep-ex]}}.

\bibitem{Wang:1991hta}
X.-N. Wang and M.~Gyulassy, ``{HIJING: A Monte Carlo model for multiple jet
  production in pp, p A and A A collisions},''
\href{http://dx.doi.org/10.1103/PhysRevD.44.3501}{{\em Phys. Rev. D} {\bfseries
  44} (1991) 3501--3516}.

\bibitem{Brun:1994aa}
R.~Brun, F.~Bruyant, F.~Carminati, S.~Giani, M.~Maire, A.~McPherson,
  G.~Patrick, and L.~Urban,
``{GEANT Detector Description and Simulation Tool},''.

\bibitem{Kopylov:1974th}
G.~I. Kopylov, ``{Like particle correlations as a tool to study the multiple
  production mechanism},''
\href{http://dx.doi.org/10.1016/0370-2693(74)90263-9}{{\em Phys. Lett. B}
  {\bfseries 50} (1974) 472--474}.

\bibitem{Brown:2005ze}
D.~A. Brown, A.~Enokizono, M.~Heffner, R.~Soltz, P.~Danielewicz, and S.~Pratt,
  ``{Imaging three dimensional two-particle correlations for heavy-ion reaction
  studies},'' \href{http://dx.doi.org/10.1103/PhysRevC.72.054902}{{\em Phys.
  Rev. C} {\bfseries 72} (2005) 054902},
\href{http://arxiv.org/abs/nucl-th/0507015}{{\ttfamily arXiv:nucl-th/0507015
  [nucl-th]}}.

\bibitem{Pratt:1986cc}
S.~Pratt, ``Pion interferometry of quark - gluon plasma,''
{\em Phys. Rev.} {\bfseries D33} (1986) 1314--1327.

\bibitem{Bertsch:1989vn}
G.~F. Bertsch, ``Pion interferometry as a probe of the plasma,''
{\em Nucl. Phys.} {\bfseries A498} (1989) 173c--180c.

\bibitem{Kisiel:2017gip}
A.~Kisiel, ``{Non-identical Particle Correlation Analysis in the Presence of
  Non-femtoscopic Correlations},''
\href{http://dx.doi.org/10.5506/APhysPolB.48.717}{{\em Acta Phys. Polon. B}
  {\bfseries 48} (2017) 717}.

\bibitem{Kisiel:2004fcn}
A.~Kisiel, ``{CorrFit - a program to fit arbitrary two-particle correlation
  functions},''
{\em Nukleonika} {\bfseries 49} no.~suppl.2, (2004) s81--s83.

\bibitem{Kisiel:2005hn}
A.~Kisiel, T.~Taluc, W.~Broniowski, and W.~Florkowski, ``{THERMINATOR: THERMal
  heavy-IoN generATOR},''
  \href{http://dx.doi.org/10.1016/j.cpc.2005.11.010}{{\em Comput. Phys.
  Commun.} {\bfseries 174} (2006) 669--687},
\href{http://arxiv.org/abs/nucl-th/0504047}{{\ttfamily arXiv:nucl-th/0504047
  [nucl-th]}}.

\end{thebibliography}\endgroup
\newpage
\appendix
\section{The ALICE Collaboration}
\label{app:collab}

\begingroup
\small
\begin{flushleft}
S.~Acharya\Irefn{org141}\And 
D.~Adamov\'{a}\Irefn{org95}\And 
A.~Adler\Irefn{org74}\And 
J.~Adolfsson\Irefn{org81}\And 
M.M.~Aggarwal\Irefn{org100}\And 
S.~Agha\Irefn{org14}\And 
G.~Aglieri Rinella\Irefn{org34}\And 
M.~Agnello\Irefn{org30}\And 
N.~Agrawal\Irefn{org10}\textsuperscript{,}\Irefn{org54}\And 
Z.~Ahammed\Irefn{org141}
% ...
\And
\renewcommand\labelenumi{\textsuperscript{\theenumi}~}

\section*{Affiliation notes}
\renewcommand\theenumi{\roman{enumi}}
\begin{Authlist}
\item \Adef{org*}Deceased
\item \Adef{orgI}Italian National Agency for New Technologies, Energy and Sustainable Economic Development (ENEA), Bologna, Italy
\item \Adef{orgII}Dipartimento DET del Politecnico di Torino, Turin, Italy
\item \Adef{orgIII}M.V. Lomonosov Moscow State University, D.V. Skobeltsyn Institute of Nuclear, Physics, Moscow, Russia
\item \Adef{orgIV}Department of Applied Physics, Aligarh Muslim University, Aligarh, India
\item \Adef{orgV}Institute of Theoretical Physics, University of Wroclaw, Poland
\end{Authlist}

\section*{Collaboration Institutes}
\renewcommand\theenumi{\arabic{enumi}~}
\begin{Authlist}
\item \Idef{org1}A.I. Alikhanyan National Science Laboratory (Yerevan Physics Institute) Foundation, Yerevan, Armenia
\item \Idef{org2}Bogolyubov Institute for Theoretical Physics, National Academy of Sciences of Ukraine, Kiev, Ukraine
\item \Idef{org3}Bose Institute, Department of Physics  and Centre for Astroparticle Physics and Space Science (CAPSS), Kolkata, India
\item \Idef{org4}Budker Institute for Nuclear Physics, Novosibirsk, Russia
\item \Idef{org5}California Polytechnic State University, San Luis Obispo, California, United States
\item \Idef{org6}Central China Normal University, Wuhan, China
\item \Idef{org7}Centre de Calcul de l'IN2P3, Villeurbanne, Lyon, France
\item \Idef{org8}Centro de Aplicaciones Tecnol\'{o}gicas y Desarrollo Nuclear (CEADEN), Havana, Cuba
\item \Idef{org9}Centro de Investigaci\'{o}n y de Estudios Avanzados (CINVESTAV), Mexico City and M\'{e}rida, Mexico
\item \Idef{org10}Centro Fermi - Museo Storico della Fisica e Centro Studi e Ricerche ``Enrico Fermi', Rome, Italy
\item \Idef{org11}Chicago State University, Chicago, Illinois, United States
\item \Idef{org12}China Institute of Atomic Energy, Beijing, China
\item \Idef{org13}Comenius University Bratislava, Faculty of Mathematics, Physics and Informatics, Bratislava, Slovakia
\item \Idef{org14}COMSATS University Islamabad, Islamabad, Pakistan
\item \Idef{org15}Creighton University, Omaha, Nebraska, United States
\item \Idef{org16}Department of Physics, Aligarh Muslim University, Aligarh, India
\item \Idef{org17}Department of Physics, Pusan National University, Pusan, Republic of Korea
\item \Idef{org18}Department of Physics, Sejong University, Seoul, Republic of Korea
\item \Idef{org19}Department of Physics, University of California, Berkeley, California, United States
\item \Idef{org20}Department of Physics, University of Oslo, Oslo, Norway
\item \Idef{org21}Department of Physics and Technology, University of Bergen, Bergen, Norway
\item \Idef{org22}Dipartimento di Fisica dell'Universit\`{a} 'La Sapienza' and Sezione INFN, Rome, Italy
\item \Idef{org23}Dipartimento di Fisica dell'Universit\`{a} and Sezione INFN, Cagliari, Italy
\item \Idef{org24}Dipartimento di Fisica dell'Universit\`{a} and Sezione INFN, Trieste, Italy
\item \Idef{org25}Dipartimento di Fisica dell'Universit\`{a} and Sezione INFN, Turin, Italy
\item \Idef{org26}Dipartimento di Fisica e Astronomia dell'Universit\`{a} and Sezione INFN, Bologna, Italy
\item \Idef{org27}Dipartimento di Fisica e Astronomia dell'Universit\`{a} and Sezione INFN, Catania, Italy
\item \Idef{org28}Dipartimento di Fisica e Astronomia dell'Universit\`{a} and Sezione INFN, Padova, Italy
\item \Idef{org29}Dipartimento di Fisica `E.R.~Caianiello' dell'Universit\`{a} and Gruppo Collegato INFN, Salerno, Italy
\item \Idef{org30}Dipartimento DISAT del Politecnico and Sezione INFN, Turin, Italy
\item \Idef{org31}Dipartimento di Scienze e Innovazione Tecnologica dell'Universit\`{a} del Piemonte Orientale and INFN Sezione di Torino, Alessandria, Italy
\item \Idef{org32}Dipartimento di Scienze MIFT, Universit\`{a} di Messina, Messina, Italy
\item \Idef{org33}Dipartimento Interateneo di Fisica `M.~Merlin' and Sezione INFN, Bari, Italy
\item \Idef{org34}European Organization for Nuclear Research (CERN), Geneva, Switzerland
\item \Idef{org35}Faculty of Electrical Engineering, Mechanical Engineering and Naval Architecture, University of Split, Split, Croatia
\item \Idef{org36}Faculty of Engineering and Science, Western Norway University of Applied Sciences, Bergen, Norway
\item \Idef{org37}Faculty of Nuclear Sciences and Physical Engineering, Czech Technical University in Prague, Prague, Czech Republic
\item \Idef{org38}Faculty of Science, P.J.~\v{S}af\'{a}rik University, Ko\v{s}ice, Slovakia
\item \Idef{org39}Frankfurt Institute for Advanced Studies, Johann Wolfgang Goethe-Universit\"{a}t Frankfurt, Frankfurt, Germany
\item \Idef{org40}Fudan University, Shanghai, China
\item \Idef{org41}Gangneung-Wonju National University, Gangneung, Republic of Korea
\item \Idef{org42}Gauhati University, Department of Physics, Guwahati, India
\item \Idef{org43}Helmholtz-Institut f\"{u}r Strahlen- und Kernphysik, Rheinische Friedrich-Wilhelms-Universit\"{a}t Bonn, Bonn, Germany
\item \Idef{org44}Helsinki Institute of Physics (HIP), Helsinki, Finland
\item \Idef{org45}High Energy Physics Group,  Universidad Aut\'{o}noma de Puebla, Puebla, Mexico
\item \Idef{org46}Hiroshima University, Hiroshima, Japan
\item \Idef{org47}Hochschule Worms, Zentrum  f\"{u}r Technologietransfer und Telekommunikation (ZTT), Worms, Germany
\item \Idef{org48}Horia Hulubei National Institute of Physics and Nuclear Engineering, Bucharest, Romania
\item \Idef{org49}Indian Institute of Technology Bombay (IIT), Mumbai, India
\item \Idef{org50}Indian Institute of Technology Indore, Indore, India
\item \Idef{org51}Indonesian Institute of Sciences, Jakarta, Indonesia
\item \Idef{org52}INFN, Laboratori Nazionali di Frascati, Frascati, Italy
\item \Idef{org53}INFN, Sezione di Bari, Bari, Italy
\item \Idef{org54}INFN, Sezione di Bologna, Bologna, Italy
\item \Idef{org55}INFN, Sezione di Cagliari, Cagliari, Italy
\item \Idef{org56}INFN, Sezione di Catania, Catania, Italy
\item \Idef{org57}INFN, Sezione di Padova, Padova, Italy
\item \Idef{org58}INFN, Sezione di Roma, Rome, Italy
\item \Idef{org59}INFN, Sezione di Torino, Turin, Italy
\item \Idef{org60}INFN, Sezione di Trieste, Trieste, Italy
\item \Idef{org61}Inha University, Incheon, Republic of Korea
\item \Idef{org62}Institute for Nuclear Research, Academy of Sciences, Moscow, Russia
\item \Idef{org63}Institute for Subatomic Physics, Utrecht University/Nikhef, Utrecht, Netherlands
\item \Idef{org64}Institute of Experimental Physics, Slovak Academy of Sciences, Ko\v{s}ice, Slovakia
\item \Idef{org65}Institute of Physics, Homi Bhabha National Institute, Bhubaneswar, India
\item \Idef{org66}Institute of Physics of the Czech Academy of Sciences, Prague, Czech Republic
\item \Idef{org67}Institute of Space Science (ISS), Bucharest, Romania
\item \Idef{org68}Institut f\"{u}r Kernphysik, Johann Wolfgang Goethe-Universit\"{a}t Frankfurt, Frankfurt, Germany
\item \Idef{org69}Instituto de Ciencias Nucleares, Universidad Nacional Aut\'{o}noma de M\'{e}xico, Mexico City, Mexico
\item \Idef{org70}Instituto de F\'{i}sica, Universidade Federal do Rio Grande do Sul (UFRGS), Porto Alegre, Brazil
\item \Idef{org71}Instituto de F\'{\i}sica, Universidad Nacional Aut\'{o}noma de M\'{e}xico, Mexico City, Mexico
\item \Idef{org72}iThemba LABS, National Research Foundation, Somerset West, South Africa
\item \Idef{org73}Jeonbuk National University, Jeonju, Republic of Korea
\item \Idef{org74}Johann-Wolfgang-Goethe Universit\"{a}t Frankfurt Institut f\"{u}r Informatik, Fachbereich Informatik und Mathematik, Frankfurt, Germany
\item \Idef{org75}Joint Institute for Nuclear Research (JINR), Dubna, Russia
\item \Idef{org76}Korea Institute of Science and Technology Information, Daejeon, Republic of Korea
\item \Idef{org77}KTO Karatay University, Konya, Turkey
\item \Idef{org78}Laboratoire de Physique des 2 Infinis, Ir\`{e}ne Joliot-Curie, Orsay, France
\item \Idef{org79}Laboratoire de Physique Subatomique et de Cosmologie, Universit\'{e} Grenoble-Alpes, CNRS-IN2P3, Grenoble, France
\item \Idef{org80}Lawrence Berkeley National Laboratory, Berkeley, California, United States
\item \Idef{org81}Lund University Department of Physics, Division of Particle Physics, Lund, Sweden
\item \Idef{org82}Nagasaki Institute of Applied Science, Nagasaki, Japan
\item \Idef{org83}Nara Women{'}s University (NWU), Nara, Japan
\item \Idef{org84}National and Kapodistrian University of Athens, School of Science, Department of Physics , Athens, Greece
\item \Idef{org85}National Centre for Nuclear Research, Warsaw, Poland
\item \Idef{org86}National Institute of Science Education and Research, Homi Bhabha National Institute, Jatni, India
\item \Idef{org87}National Nuclear Research Center, Baku, Azerbaijan
\item \Idef{org88}National Research Centre Kurchatov Institute, Moscow, Russia
\item \Idef{org89}Niels Bohr Institute, University of Copenhagen, Copenhagen, Denmark
\item \Idef{org90}Nikhef, National institute for subatomic physics, Amsterdam, Netherlands
\item \Idef{org91}NRC Kurchatov Institute IHEP, Protvino, Russia
\item \Idef{org92}NRC \guillemotleft Kurchatov\guillemotright  Institute - ITEP, Moscow, Russia
\item \Idef{org93}NRNU Moscow Engineering Physics Institute, Moscow, Russia
\item \Idef{org94}Nuclear Physics Group, STFC Daresbury Laboratory, Daresbury, United Kingdom
\item \Idef{org95}Nuclear Physics Institute of the Czech Academy of Sciences, \v{R}e\v{z} u Prahy, Czech Republic
\item \Idef{org96}Oak Ridge National Laboratory, Oak Ridge, Tennessee, United States
\item \Idef{org97}Ohio State University, Columbus, Ohio, United States
\item \Idef{org98}Petersburg Nuclear Physics Institute, Gatchina, Russia
\item \Idef{org99}Physics department, Faculty of science, University of Zagreb, Zagreb, Croatia
\item \Idef{org100}Physics Department, Panjab University, Chandigarh, India
\item \Idef{org101}Physics Department, University of Jammu, Jammu, India
\item \Idef{org102}Physics Department, University of Rajasthan, Jaipur, India
\item \Idef{org103}Physikalisches Institut, Eberhard-Karls-Universit\"{a}t T\"{u}bingen, T\"{u}bingen, Germany
\item \Idef{org104}Physikalisches Institut, Ruprecht-Karls-Universit\"{a}t Heidelberg, Heidelberg, Germany
\item \Idef{org105}Physik Department, Technische Universit\"{a}t M\"{u}nchen, Munich, Germany
\item \Idef{org106}Politecnico di Bari, Bari, Italy
\item \Idef{org107}Research Division and ExtreMe Matter Institute EMMI, GSI Helmholtzzentrum f\"ur Schwerionenforschung GmbH, Darmstadt, Germany
\item \Idef{org108}Rudjer Bo\v{s}kovi\'{c} Institute, Zagreb, Croatia
\item \Idef{org109}Russian Federal Nuclear Center (VNIIEF), Sarov, Russia
\item \Idef{org110}Saha Institute of Nuclear Physics, Homi Bhabha National Institute, Kolkata, India
\item \Idef{org111}School of Physics and Astronomy, University of Birmingham, Birmingham, United Kingdom
\item \Idef{org112}Secci\'{o}n F\'{\i}sica, Departamento de Ciencias, Pontificia Universidad Cat\'{o}lica del Per\'{u}, Lima, Peru
\item \Idef{org113}St. Petersburg State University, St. Petersburg, Russia
\item \Idef{org114}Stefan Meyer Institut f\"{u}r Subatomare Physik (SMI), Vienna, Austria
\item \Idef{org115}SUBATECH, IMT Atlantique, Universit\'{e} de Nantes, CNRS-IN2P3, Nantes, France
\item \Idef{org116}Suranaree University of Technology, Nakhon Ratchasima, Thailand
\item \Idef{org117}Technical University of Ko\v{s}ice, Ko\v{s}ice, Slovakia
\item \Idef{org118}The Henryk Niewodniczanski Institute of Nuclear Physics, Polish Academy of Sciences, Cracow, Poland
\item \Idef{org119}The University of Texas at Austin, Austin, Texas, United States
\item \Idef{org120}Universidad Aut\'{o}noma de Sinaloa, Culiac\'{a}n, Mexico
\item \Idef{org121}Universidade de S\~{a}o Paulo (USP), S\~{a}o Paulo, Brazil
\item \Idef{org122}Universidade Estadual de Campinas (UNICAMP), Campinas, Brazil
\item \Idef{org123}Universidade Federal do ABC, Santo Andre, Brazil
\item \Idef{org124}University of Cape Town, Cape Town, South Africa
\item \Idef{org125}University of Houston, Houston, Texas, United States
\item \Idef{org126}University of Jyv\"{a}skyl\"{a}, Jyv\"{a}skyl\"{a}, Finland
\item \Idef{org127}University of Liverpool, Liverpool, United Kingdom
\item \Idef{org128}University of Science and Technology of China, Hefei, China
\item \Idef{org129}University of South-Eastern Norway, Tonsberg, Norway
\item \Idef{org130}University of Tennessee, Knoxville, Tennessee, United States
\item \Idef{org131}University of the Witwatersrand, Johannesburg, South Africa
\item \Idef{org132}University of Tokyo, Tokyo, Japan
\item \Idef{org133}University of Tsukuba, Tsukuba, Japan
\item \Idef{org134}Universit\'{e} Clermont Auvergne, CNRS/IN2P3, LPC, Clermont-Ferrand, France
\item \Idef{org135}Universit\'{e} de Lyon, Universit\'{e} Lyon 1, CNRS/IN2P3, IPN-Lyon, Villeurbanne, Lyon, France
\item \Idef{org136}Universit\'{e} de Strasbourg, CNRS, IPHC UMR 7178, F-67000 Strasbourg, France, Strasbourg, France
\item \Idef{org137}Universit\'{e} Paris-Saclay Centre d'Etudes de Saclay (CEA), IRFU, D\'{e}partment de Physique Nucl\'{e}aire (DPhN), Saclay, France
\item \Idef{org138}Universit\`{a} degli Studi di Foggia, Foggia, Italy
\item \Idef{org139}Universit\`{a} degli Studi di Pavia, Pavia, Italy
\item \Idef{org140}Universit\`{a} di Brescia, Brescia, Italy
\item \Idef{org141}Variable Energy Cyclotron Centre, Homi Bhabha National Institute, Kolkata, India
\item \Idef{org142}Warsaw University of Technology, Warsaw, Poland
\item \Idef{org143}Wayne State University, Detroit, Michigan, United States
\item \Idef{org144}Westf\"{a}lische Wilhelms-Universit\"{a}t M\"{u}nster, Institut f\"{u}r Kernphysik, M\"{u}nster, Germany
\item \Idef{org145}Wigner Research Centre for Physics, Budapest, Hungary
\item \Idef{org146}Yale University, New Haven, Connecticut, United States
\item \Idef{org147}Yonsei University, Seoul, Republic of Korea
\end{Authlist}
\endgroup
%
%
\end{document}